\documentclass[showpacs,floatfix,preprint]{revtex4-1}
\usepackage{graphicx,psfrag,amsmath,amssymb,amsfonts,latexsym,color,epsf,graphpap,dcolumn}

\date{today}
\usepackage[latin1]{inputenc}
\usepackage[T1]{fontenc} 
\usepackage[english]{babel}
\usepackage{amsmath}
\usepackage{amsfonts} 
\usepackage{amssymb} 
\usepackage{graphicx}  
\usepackage{bm}       
\usepackage{hyperref}
\pdfoutput=1

\begin{document}
\title{Motion-induced radiation due to an atom in the presence of
a graphene plane}

\author{C\'esar D. Fosco $^1$ \footnote{fosco@cab.cnea.gov.ar}}
\author{Fernando C. Lombardo$^2$ \footnote{lombardo@df.uba.ar}}
\author{Francisco D. Mazzitelli$^1$ \footnote{fdmazzi@cab.cnea.gov.ar}}

\affiliation{$^1$ Centro At\'omico Bariloche and Instituto Balseiro,
Comisi\'on Nacional de Energ\'\i a At\'omica, 
R8402AGP Bariloche, Argentina}
\affiliation{$^2$ Departamento de F\'\i sica {\it Juan Jos\'e
 Giambiagi}, FCEyN UBA, Facultad de Ciencias Exactas y Naturales,
 Ciudad Universitaria, Pabell\' on I, 1428 Buenos Aires, Argentina }
\date{\today}
\begin{abstract}
We study the motion-induced radiation due to the non-relativistic motion of
an atom, coupled to the vacuum electromagnetic field by an electric dipole
term, in the presence of a static graphene plate. After computing the
probability of emission for an accelerated atom in empty space, we evaluate
the corrections due to the presence of the plate. We show that the
effect of the plate is to increase the probability of emission when the
atom is near the plate and oscillates along a direction perpendicular to
it. On the contrary, for parallel oscillations there is a suppression. 

\noindent We also evaluate the quantum friction on an atom moving at
constant velocity parallel to the plate. We show that there is a
threshold for quantum friction: friction occurs only when the velocity
of the atom is larger than the Fermi velocity of the electrons in
graphene.  
\end{abstract} 
\maketitle
\section{Introduction}\label{sec:intro}
Casimir and Casimir-Polder forces are physical manifestations of the vacuum
fluctuations of the electromagnetic field which involve, respectively, the
interaction between static bodies, or between an atom and a body.  The
dependence of those effects on the geometry of the system, as well as on
the electromagnetic properties of the material media and the atom, has been
intensively investigated in the last decades, both at zero and nonzero
temperatures~\cite{books}.

Graphene, a single layer of carbon atoms, can be effectively described as a
two-dimensional material. It owes its remarkable physical properties to its
planar hexagonal crystal structure, and to the fact that the electronic
degrees of freedom can be described, at low energies, as Dirac fermions:
indeed, they satisfy a linear dispersion relation, i.e, they behave as
massless fermions that propagate with the Fermi velocity $v_F\simeq 0.003 c$ \cite{graphene}. 
This yields an unusual behavior for the conductivity, as well as peculiar transport and optical
properties~\cite{graphprop}. These features have understandably raised the interest
in the analysis of the interaction of graphene with the vacuum electromagnetic
field fluctuations~\cite{graphCas}. 

In natural units, the response function of graphene is determined by two
dimensionless quantities, $v_F$, and the fine structure constant.
Therefore, in the simplest configuration of two planar graphene sheets
separated by a distance $d$ at zero temperature, simple dimensional
analysis implies that the static Casimir force has the same distance
dependence as for perfect conductors; namely, it is proportional to
$1/d^{4}$ (the force is weaker than for ideal conductors).  However, new
interesting effects appear at finite temperature~\cite{graphCasT}.
Besides, the phenomenology becomes richer when considering more realistic
descriptions for graphene; for example by including a gap in the dispersion
relation or a  non-vanishing chemical potential~\cite{graphnonideal, graphreview}. 

From a theoretical standpoint, the system is often amenable to a full {\it
ab-initio} description in the context of a continuum quantum field theory,
treating the microscopic degrees of freedom  as Dirac fields in two
dimensions~\cite{abinitio}. For the alluded Casimir and Casimir-Polder
forces, the results of this approach have been shown to be equivalent to
the ones derived from Lifshitz theory~\cite{graphreview}.

When atoms or macroscopic bodies are set into motion,  new
phenomena may appear.  Motion induced radiation, or dynamical Casimir effect
(DCE), consists in the production of photons from the vacuum due to
acceleration~\cite{reviewDCE}. A related but somewhat different effect is
quantum friction (QF), which manifests itself on objects with a constant relative sliding
velocity. Its origin is in the excitation of microscopic degrees of
freedom of the bodies, mediated by the electromagnetic field. This results
in a contactless dissipative force~\cite{qfriction}. The interplay between atomic motion and quantum vacuum fluctuations has been analyzed in different
contexts \cite{interplay}.

The DCE has been investigated in different models and physical setups,
starting from the simplest case of a one-dimensional field theory in a
cavity with a moving boundary~\cite{Moore}; then a single accelerated
mirror~\cite{Fulling}, and more recent calculations which focus on resonant
effects~\cite{reviewDCE}. The microscopic origin of the
DCE~\cite{Paulo2018, Belen2019} can be traced back to photon emission
processes in accelerated atoms: indeed, even when the oscillation frequency
is not sufficient to excite the atom, the coupling to the quantum
electromagnetic field allows for the emission of photon pairs.  Production
of photons out of the vacuum can also occur in situations where the
electromagnetic properties of a media change with time~\cite{dcemat}. A
practical implementation of this, involving superconducting circuits, has
led to the first experimental observation of the DCE~\cite{dceexp}. 

QF has also been discussed in a variety of situations,
most of them involving planar sheets or semi-spaces electromagnetically
described in terms of their dielectric properties. Due to its  short range 
and very small magnitude, QF has eluded detection so far. In spite of this,
there could be observable traces of this effect in
the velocity dependence of corrections to the accumulated
geometric phase of a neutral particle which moves with constant velocity in
front of an imperfect mirror~\cite{Lomb2020}.  It has also been shown that QF
could influence the coherences of a two-level atom~\cite{Lomb2021}.
Furthermore, an innovative experiment was designed to track traces of QF by
measuring this corrections to the geometric phase~\cite{qute21}. This 
experimentally viable scheme can spark, we believe, hope for the
detection of non-contact friction.  

Within the previous context of quantum dissipative effects, in this paper,
we study a particular system which consists of an (externally driven)
atom which moves non-relativistically (with or without acceleration) in the
presence of a planar, static graphene sheet.  
Both systems are coupled to the vacuum electromagnetic (EM) field, which
mediates the correlation between the quantum fluctuations of the charged
degrees of freedom, located in the atom and on the graphene sheet. The atom
is coupled to the EM field through its electric dipole moment. 
Because of the special features of graphene, for example, the dependence of
its response function on just two dimensionless parameters, we expect an
interesting manifestation of microscopic (i.e., one of the bodies involved
is an atom) DCE and QF.
  
In a previous work~\cite{Belen2017}, we have studied QF between two sliding
graphene sheets, and found that there was a velocity threshold for the
occurrence of friction, given by the Fermi velocity $v_F$. This fact could
be eventually relevant, in future technological applications, in order to
avoid dissipative effects. Note, however, that by the same token it  makes
QF almost impossible to detect in that kind of system. 
As we shall see, a similar threshold appears for the case of an atom moving
with constant velocity over a graphene sheet; we note in pass that the
threshold may be less difficult to overcome for the motion of a single atom
than for the collective motion of a whole plate. 
  
We have also previously studied the microscopic DCE due to an atom, with an
internal structure modelled by a quantum harmonic oscillator, coupled to a
scalar field, in the presence of an imperfect mirror~\cite{Belen2019}. The
internal degrees of freedom of the mirror have been described in terms of a set of
harmonic oscillators. We have found an interesting effect when the
mechanical frequency equals to the sum of the internal frequency of the
atom and that of the microscopic oscillators, with regions of
enhancement and suppression of the vacuum persistence amplitude. This
behavior is similar to that of spontaneous emission for an atom immersed
into a dielectric~\cite{spontemis}. We expect a rather different behaviour
in the presence of the graphene plate, due to the already mentioned absence of a parameter
with dimensions in the response function.

Many tools can be used to study the existence  and magnitude of dissipative
phenomena in this context; among them is the imaginary part of the in-out
effective action, related with the vacuum persistence amplitude, and obtained by integrating out the quantum degrees of
freedom. This results in an imaginary part which is a functional of the trajectory of the atom,
the object that we evaluate here. 

This paper is organised as follows: in Sect.~\ref{sec:themodel}, we define
the model at a microscopic level. Then, in Sect.~\ref{sec:gammaeff}, we
integrate out the charged degrees of freedom, to obtain the effective
action for the gauge field,  which also depends functionally on the
trajectory of the atom.  Results for the remaining functional integral
(i.e., over the gauge field) are presented in Sect.~\ref{sec:pert}, in a
perturbative approach.  To first order, the imaginary part of the effective
action gives the probability of emission of the atom oscillating in vacuum,
that, up to this order, and to the quadratic order in the oscillation
amplitude, is non-vanishing only when the frequency of
oscillation is larger than the  frequency of the harmonic oscillator that
describes the atom. At higher orders in the amplitude, the threshold is
reduced.

The second order contribution contains information about the influence of
the presence of the graphene sheet on the empty space vacuum persistence
amplitude. In Sect.~\ref{sec:results}, we present results about the
imaginary part of the in-out effective action for different kinds of
motion. We first consider the QF on an atom moving with constant velocity,
parallel to the plane, and show that there is a threshold for QF.  Then we
analyze the case of an oscillatory motion, perpendicular or parallel to the
plane. We find that the effect of the plane on the emission probability
depends crucially on the direction of motion. In Sect.~\ref{sec:exact} we
comment on the changes in the expression for the effective action when
the exact propagator for the gauge field in the presence of the graphene
plane is used, and the way to implement the perfect conductor limit. 
Sect.~\ref{sec:conc} contains the conclusions of our work.
\section{The microscopic model}\label{sec:themodel}
  We begin by defining the model in terms of its action ${\mathcal
S}$, depending on the intervening degrees of freedom:
$A_\mu$, a $4$-potential corresponding to the EM field, 
a Dirac field $\psi,\, \bar{\psi}$ in $2+1$ dimensions, for the electronic
degrees of freedom on the graphene sheet, ${\mathbf r}(t)$,
the center of mass of the atom, and ${\mathbf x}(t)$, the position of the
electron (relative to that center of mass). 

${\mathcal S}$ has the structure: 
\begin{equation}
{\mathcal S} \,=\, {\mathcal S}(A; {\mathbf r}, {\mathbf x};
\bar{\psi}, \psi) \,=\, {\mathcal S}_0(A) \,+\, {\mathcal S}_g(\bar{\psi},\psi; A) 
\,+\, {\mathcal S}_a({\mathbf r}, {\mathbf x};A) \;,
\end{equation}
where ${\mathcal S}_0(A)$ denotes the free EM action, while 
${\mathcal S}_g$ and ${\mathcal S}_a$ are the graphene and atom actions, respectively, each one 
including its coupling to the gauge field.

The free EM field action ${\mathcal S}_0(A)$, including a gauge fixing
term, is given by
\begin{equation}\label{eq:defsg0}
	{\mathcal S}_0(A) \;=\; \int d^4x \, 
	\left[-\frac{1}{4}  F_{\mu\nu} F^{\mu\nu} \,-\, \frac{\lambda}{2}
	(\partial \cdot A)^2 \right] \;,
\end{equation}
with $F_{\mu\nu} = \partial_\mu A_\nu - \partial_\nu A_\mu$. 
Indices from the middle of the Greek alphabet ($\mu, \nu, \ldots$) run from
$0$ to $3$, with $x^0 \equiv c t$. In our conventions, $c \equiv 1$,
and we use the metric signature $(+---)$.

The graphene action, ${\mathcal S}_g$, on the other hand, is localized on
the region occupied by the sheet which, in our choice of coordinates,
corresponds to $x^3=0$. Fields restricted to such region will therefore
depend on the reduced, $2+1$-dimensional space-time coordinates
$(x^0,x^1,x^2)$, which we denote collectively as $x_\shortparallel$. For an
unstrained plate, the form of the action for a single fermionic flavour
becomes: 
\begin{equation}
	{\mathcal S}_g(\bar{\psi},\psi; A) \;=\; \int d^3x \,
	\bar{\psi}(x_\shortparallel)  (i \rho^\alpha_\beta \gamma^\beta
	D_\alpha - m) {\psi}(x_\shortparallel) \;,    
\end{equation}
where \mbox{$D_\alpha = \partial_\alpha + i e
A_\alpha(x_\shortparallel,0)$}. Indices of the beginning of the Greek
alphabet ($\alpha, \, \beta, \ldots$) run over the values $0$, $1$, and
$2$, and $(\rho^\alpha_\beta) = {\rm diag}(1,v_F,v_F)$, with $v_F$ the
Fermi velocity. A single flavour corresponds to two $2$-component spinors,
and the representation chosen for the Dirac's $\gamma$-matrices is such
that parity is preserved even when the mass $m \neq 0$. 

The atom, on the other hand, is described by a simplified model in which a single electron of
charge $e$ and mass $m$ moves in the presence of a central potential $V$
with origin at the nucleus, where most of the mass $M$ of the atom, and a
charge $-e$, are located. 

The trajectory of the center of mass of the atom is practically
identical to the one of the nucleus, and we describe it by the vector
function $t \, \to \, {\mathbf r}(t)$.  
Denoting by ${\mathbf x}(t)$ the position of the electron with respect to
the nucleus, the interaction action between the atom and the EM
field is:
\begin{align}
{\mathcal S}_a^{(int)} \;=\; \int dt \, \big[ & e \big( \dot{\mathbf r}(t) + \dot{\mathbf
x}(t) \big) \cdot {\mathbf A}(t, {\mathbf r}(t) + {\mathbf x}(t))
\,-\, e \, A_0(t,{\mathbf r}(t)  + {\mathbf x}(t)) \nonumber\\
&-\; e  \dot{\mathbf r}(t) \cdot {\mathbf A}(t, {\mathbf r}(t))
	\,+\, e \, A_0(t,{\mathbf r}(t) ) \big]\;.
\end{align}
In the dipole approximation, i.e., assuming the fields vary smoothly on the
spatial region where the electron's wavefunction is spatially concentrated, we expand the interaction action as 

\begin{equation}
{\mathcal S}_a^{(int)} \approx \; e \int dt \, \left[ \dot{ {\mathbf r}} \cdot \left({\mathbf A}(t, {\mathbf r} + {\mathbf x}) -   {\mathbf A}(t, {\mathbf r}) \right) 
+ {\dot  {\mathbf x}} \cdot {\mathbf A}(t, {\mathbf r} + {\mathbf x}) -  {\mathbf x} \cdot  \nabla \phi(t, {\mathbf r} )  \right].  \nonumber \end{equation}  
After making an integration by parts and also using that 
$|\dot{\mathbf r}| << |\dot{\mathbf x}|$, last expression can be written as 
\begin{equation}
{\mathcal S}_a^{(int)} \approx  \int dt \, \left[ {\mathbf d}(t) \cdot {\mathbf E}(t,{\mathbf r}(t)) + {\mathbf m}(t) \cdot {\mathbf B}(t,{\mathbf r}(t)) \right], \nonumber
\end{equation}
with ${\mathbf d}$ the electric dipole moment ${\mathbf d}(t) = e {\mathbf x}(t)$,  and ${\mathbf m}$ the magnetic dipole  ${\mathbf m} = e/2 \; ({\mathbf x}(t) \times {\dot {\mathbf x}}(t))$. 
Neglecting the magnetic dipole term, one gets ${\mathcal S}_a^{(int)} 
\simeq  {\mathcal S}_a^{(dip)}$, where:
\begin{equation}
	{\mathcal S}_a^{(dip)} \;=\; \int dt \; {\mathbf d}(t) \cdot {\mathbf E}(t,{\mathbf
	r}(t)) \;.
\end{equation}.

The total action for the atom, 	${\mathcal S}_a$, consistent with the
approximations above, is then:
\begin{equation}
{\mathcal S}_a \;=\; \int dt \, 
	\Big[ \frac{m}{2} \dot{\mathbf x}^2(t) \,-\, V({\mathbf x}(t))
\,+\, {\mathbf d}(t) \cdot {\mathbf E}(t,{\mathbf r}(t)) \Big]\;,
\end{equation}
where we have ignored kinetic and potential terms for ${\mathbf r}$, since
its dynamics is externally determined.

\section{Effective action}\label{sec:gammaeff}
The effective action $\Gamma[{\mathbf r}(t)]$ is obtained by
functional integration of all the degrees of freedom, with Feynman's $i
\epsilon$ prescription implicitly assumed for all the integrals, since
we are interested in the in-out effective action. Namely \cite{FLM2007},
\begin{equation}
	e^{i \Gamma({\mathbf r})} \;=\; 
\frac{\int {\mathcal D}A \, {\mathcal D}\bar{\psi} {\mathcal D}\psi \, 
{\mathcal D}{\mathbf x} \, e^{ i {\mathcal S}(A; {\mathbf r}, {\mathbf x};
\bar{\psi}, \psi)}}{\int {\mathcal D}A \, {\mathcal D}\bar{\psi} {\mathcal D}\psi \, 
{\mathcal D}{\mathbf x} \, e^{ i {\mathcal S}(A; {\mathbf r}_0, {\mathbf x};
	\bar{\psi}, \psi)}}
\;,
\end{equation}
where ${\mathbf r}_0$ is some `reference' trajectory. We will, in most
of the applications, use  a time-independent ${\mathbf r}_0$. This is
useful when considering a bounded motion, where it is natural to identify
it with the time average of ${\mathbf r}(t)$ .

We find it useful to decompose the integral into two successive steps, 
\begin{align}
e^{i \Gamma({\mathbf r})} &=\; 
\frac{\int {\mathcal D}A \, e^{ i {\mathcal S}_{\rm eff}(A; {\mathbf
	r})}}{\int {\mathcal D}A \, e^{ i {\mathcal S}_{\rm eff}(A; {\mathbf
	r}_0)}}
	\;\;, \;\;\; \; e^{i {\mathcal S}_{\rm eff}(A; {\mathbf r})} \;=\;  \int
	{\mathcal D}\bar{\psi} {\mathcal D}\psi \, 
{\mathcal D}{\mathbf x} \, 
e^{ i {\mathcal S}(A; {\mathbf r}, {\mathbf x}; \bar{\psi}, \psi)}
\end{align}
where ${\mathcal S}_{\rm eff}(A; {\mathbf r})$ is assumed to be defined
modulo an irrelevant constant, which cancels out in $\Gamma$.
It is convenient to set:
${\mathcal S}_{\rm eff}(A; {\mathbf r}) = {\mathcal S}_0(A) + {\mathcal
S}_{\rm I}(A; {\mathbf r})$ 
where we have separated from ${\mathcal S}_{\rm eff}(A;{\mathbf r})$ the free
EM field contribution from the one due to the charged sources.

Therefore, we need now to integrate the fermionic field and the electron
trajectories. In general, neither of them can be performed exactly, so,
consistently with the assumptions we have already mentioned, we will retain the
terms up to  order $e^2$. Namely, both will produce terms which are quadratic in  $A$. 

Let us find the explicit form of those terms. In an obvious notation: 
\begin{equation}
{\mathcal S}_{\rm I}(A; {\mathbf r}) 
\;=\; {\mathcal S}_{\rm I}^{(a)}(A; {\mathbf r}) 
\,+\,{\mathcal S}_{\rm I}^{(g)}(A) \;.
\end{equation}
To find ${\mathcal S}_{\rm I}^{(a)}$, one needs to perform the
functional integral 
\begin{equation}\label{eq:wj}
	e^{i{\mathcal  S}_{\rm I}^{(a)}(A; {\mathbf r})} \;=\; {\mathcal{N}}
\int {\mathcal D}{\mathbf x} \, e^{ i \int dt \, 
	\big( \frac{m}{2} \dot{\mathbf x}^2(t) \,-\, V({\mathbf x} (t))
	\,+\, e {\mathbf x}(t) \cdot {\mathbf E}(t,{\mathbf r}(t))
	\big)}\;,
\end{equation}
which, as usual, is  assumed to vanish in the noninteracting case, i.e.  ${\mathcal  S}_{\rm
I}^{(a)}\Big|_{e=0} = 0$. This can be achieved by a proper choice  of the constant $\mathcal N$.

We recognise the structure of ${\mathcal  S}_{\rm I}^{(a)}(A; {\mathbf r})$
in (\ref{eq:wj}) as the generating functional
of ${\mathbf x}(t)$ connected correlation functions, regarding $e {\mathbf E}$ as the
external source. A contribution of $n^{th}$-order in a series expansion of  ${\mathcal
S}_{\rm I}^{(a)}$ in powers of $e$, will therefore be determined by 
the corresponding $n$-leg connected correlation function of ${\mathbf x}(t)$, in
the absence of any external source.
In particular, to the second order in the source,
\begin{equation}
	{\mathcal  S}_{\rm I}^{(a)}(A; {\mathbf r}) \;=\; \frac{i e^2}{2}
	\; \int dt \int dt' \; 
	E_i(t,{\mathbf r}(t)) \, \langle x_i(t) x_j(t') \rangle \,
	E_j(t',{\mathbf r}(t')) \;,
\end{equation}
where:
\begin{equation}
	\langle x_i(t) x_j(t') \rangle \;=\; 
	\frac{\int {\mathcal D}{\mathbf x} \; x_i(t) x_j(t') \; e^{ i \int dt \, 
	\big( \frac{m}{2} \dot{\mathbf x}^2(t) \,-\, V({\mathbf x}(t))\big)}}{\int 
	{\mathcal D}{\mathbf x} \; e^{ i \int dt \, 
	\big( \frac{m}{2} \dot{\mathbf x}^2(t) \,-\, V({\mathbf x}
	(t))\big)}} \;.
\end{equation}
That correlation function, due to time-translation and rotational
invariances, regardless of the specific form of $V$, may be written as follows: 
\begin{equation}
	\langle x_i(t) x_j(t') \rangle \;=\; \delta_{ij} \; \int \frac{d\nu}{2\pi} e^{-i \nu (t-t')} \,
	\widetilde{\Delta}(\nu) \;, 
\end{equation}
with the precise form of $\widetilde{\Delta}_a(\nu)$ to be determined by
the potential. For example, for a harmonic potential of frequency $\Omega$,
$V = \frac{m}{2} \Omega^2 {\mathbf x}^2$, one has the exact result: 
\begin{equation}
	\widetilde{\Delta}(\nu) \;\equiv\; \widetilde{\Delta}_\Omega(\nu)
	\;=\; \frac{1}{m} \, \frac{i}{\nu^2 - \Omega^2 + i
	\epsilon} \;.
\end{equation}
For the case of a general potential, it is interesting to note that,
since  $\langle x_i(t) x_j(t') \rangle$ is the (exact) Feynman propagator
of the coordinate operator, we can always invoke the spectral decomposition
theorem to write:
\begin{equation}
	\widetilde{\Delta}(\nu) \;=\; \frac{1}{m} \, \int_0^\infty
	d\omega^2 \; \rho(\omega^2) \; \frac{i}{\nu^2 - \omega^2 + i \epsilon} 
\;\equiv\; \int_0^\infty
	d\omega^2 \; \rho(\omega^2) \;\widetilde{\Delta}_\omega(\nu) \;.
\end{equation}
Therefore, for a general potential $V$,
\begin{equation}
	{\mathcal  S}_{\rm I}^{(a)}(A; {\mathbf r}) \;=\; \frac{i e^2}{2}
	\int_0^\infty d\omega^2 \; \rho(\omega^2) 
	\; \int dt \int dt' \; 
	E_i(t,{\mathbf r}(t)) \, \Delta_\omega(t-t') \,
	E_i(t',{\mathbf r}(t')) \;,
\end{equation}
where $\Delta_\omega(t)$ is the inverse Fourier transform of $\widetilde{\Delta}_\omega(\nu)$.

It is noteworthy that we obtain qualitatively similar results, to the
quadratic order in the charge, if we assume that atom is described by
a two-level system. Indeed, to that order, we only need to know the
correlation function:
\begin{equation}
\langle x_i(t) x_j(t') \rangle \;=\; \langle 0 | T[\hat{x}_i(t)
\hat{x}_j(t') ] |0 \rangle \;,
\end{equation}
where $T$ denotes the time-ordered product, and $|0\rangle$ the ground
state of the atom. If now we assume that we have a two-level system, we may
produce a more explicit expression for the correlation function. Indeed, we
may insert an identity operator $I$ in the time-ordered product, written in
terms of the two eigenstates of the atom's Hamiltonian. One of those
eigenstates is of course $|0\rangle$ with energy $E_0$, and there is
just one an excited level, $|1\rangle$. For a rotationally symmetric
potential, the excited state must have a degeneracy consistent with the
conservation of angular momentum.  The simplest non-trivial one is $3$.
Therefore, we assume that the excited level has an energy $E_1$, with the
three degenerate eigenstates $|1^{(\alpha)}\rangle$, with $\alpha=1, 2, 3$.

We then insert the identity written as: $I = |0\rangle \langle 0| + \sum_{\alpha=1}^3
|1^{(\alpha)}\rangle \langle 1^{(\alpha)}|$ into the time-ordered product
of Heisenberg operators: $\hat{x}_i(t) = e^{i t \hat{H}} \hat{x}_i e^{-i t
\hat{H}}$; it is important to note that Wigner-Eckart's theorem implies
that $\hat{x}_i$ can only have non-vanishing matrix elements
between the ground state and the excited ones. Thus
\begin{equation}
\langle 0 | T[\hat{x}_i(t) \hat{x}_j(t') ] |0 \rangle \;=\;
	\theta(t-t') \, e^{-i (t-t') \Omega} \,
	\xi_{ij}
+\, \theta(t'-t) \, e^{-i (t'-t) \Omega} \,
	\xi_{ji}
\end{equation}
where $\Omega = E_1-E_0$, and
\begin{equation}
\xi_{ij} \;=\; \sum_\alpha \langle 0 |\hat{x}_i|1^{(\alpha)}\rangle
\langle 1^{(\alpha)}|\hat{x}_j|0\rangle \;.
\end{equation}
One sees that $\xi_{ij}$ behaves as a second-order tensor under rotations,
and because of the rotational symmetry of the system, can only be
proportional to $\delta_{ij}$: $\xi_{ij} \;=\; \chi^2 \,\delta_{ij}$, for some constant $\chi$. Thus,
\begin{align}
	\langle x_i(t) x_j(t) \rangle &=\;  (2 \Omega \, \chi^2 m)  \, 
	\delta_{ij} \, \frac{1}{2 m \Omega} [\theta(t-t') \, e^{-i (t-t')
	\Omega} \,
	 +\, \theta(t'-t) \, e^{-i (t'-t) \Omega} ] \nonumber\\
	 &=\;  (2 \Omega \, \chi^2 m)  \, 
	\delta_{ij} \, \Delta_\Omega(t-t') \;.
\end{align}
We conclude that, for this quadratic approximation to the effective action
due to the atom, a two-level system produces essentially the same effect as a harmonic
potential of frequency $\Omega$, if $E_1 - E_0 \,\propto\, ( 2 \chi^2 m)^{-1}$.

Regarding the graphene contribution, we see that:
\begin{equation}
{\mathcal S}_{\rm I}^{(g)}(A) \;=\; 
\frac{1}{2} \,\int d^3x_\parallel \int d^3y_\parallel \,
A^\alpha(x_\parallel,0) \, \Pi_{\alpha\beta}(x_\parallel,y_\parallel) \,
A^\beta(y_\parallel,0) \;,
\end{equation}
with the tensor kernel $\Pi_{\alpha\beta}$ denoting the vacuum polarization tensor
for the Dirac field on the plane.  

Using a tilde to denote Fourier transformation,
$\widetilde{\Pi}_{\alpha\beta}$ may be decomposed into two
irreducible tensors (projectors) adapted to the symmetries of the system.
Introducing first the ingredients: \mbox{$\breve{k}^\alpha \equiv k^\alpha - k^0 n^\alpha$},
\mbox{$\breve{g}_{\alpha\beta} \equiv g_{\alpha\beta} - n_\alpha
n_\beta$} and  \mbox{$n^\alpha = (1,0,0)$}, a convenient pair of projectors
is: ${\mathcal P}^l_{\alpha\beta}$ and ${\mathcal P}^t_{\alpha\beta}$, where:
\begin{equation}
{\mathcal P}^t_{\alpha\beta} \,\equiv\, \breve{g}_{\alpha\beta} -
\frac{\breve{k}_\alpha\breve{k}_\beta}{\breve{k}^2} \;\;,\;\;\;\;
{\mathcal P}^l_{\alpha\beta} \,\equiv\, {\mathcal P}^\perp_{\alpha\beta}\,-\, 
{\mathcal P}^t_{\alpha\beta}
\;\;,\;\;\;\;
{\mathcal P}^\perp_{\alpha\beta} \,\equiv\, g_{\alpha\beta}- \frac{k_\alpha
k_\beta}{k^2} \;.
\end{equation}

The result, for the most relevant case of $m=0$, is as follows: 
\begin{align}
	\widetilde{\Pi}_{\alpha\beta} &= \alpha_N  \, 
	\sqrt{v_F^2{\mathbf k_\shortparallel}^2 - k_0^2} \; 
	\Big[ {\mathcal P}^t_{\alpha\beta} \,+\, 
	\frac{{\mathbf k_\shortparallel}^2 - k_0^2}{v_F^2 {\mathbf
	k_\shortparallel}^2-k_0^2} {\mathcal P}^l_{\alpha\beta} \Big] \nonumber\\
	  &= \alpha_N \, \sqrt{{\mathbf k_\shortparallel}^2 - k_0^2} \,  
	\Big[ \, \sqrt{\frac{v_F^2 {\mathbf
	k_\shortparallel}^2 -k_0^2}{{\mathbf k_\shortparallel}^2 -
	k_0^2}} \;
	{\mathcal P}^t_{\alpha\beta}  + 
	\sqrt{\frac{{\mathbf k_\shortparallel}^2 - k_0^2}{v_F^2
	{\mathbf k_\shortparallel}^2-k_0^2}} \; {\mathcal P}^l_{\alpha\beta} 
	\Big] \;,
\end{align}
where we introduced
$\alpha_N \equiv \frac{e^2 N}{16}$, with $N$ the number of 2-component
Dirac fermion fields.
We have thus found the explicit form of the two terms into which ${\mathcal S}_{\rm
I}$ may be decomposed. 

It is worth noting that  assuming gauge, translation, and rotation
invariances, one can generalize the previous expression to other planar
media as follows: 
\begin{equation}\label{eq:piab}
	\widetilde{\Pi}_{\alpha\beta}(k_\shortparallel) \;=\;  g_t(k_\shortparallel) \, {\mathcal P}^t_{\alpha\beta} 
\;+\; g_l(k_\shortparallel)  \,  {\mathcal P}^l_{\alpha\beta} \;,
\end{equation}
where $g_t$ and $g_l$ are functions of $k_0$ and $|{\mathbf
k_\shortparallel}|$.

\section{Pertubative expansion}\label{sec:pert} 
We now consider the evaluation of the effective action, by proceeding to
integrate the vacuum field, in a perturbative expansion approach. 

$\Gamma[{\mathbf r}(t)]$ is given by:
\begin{equation}
	e^{i\Gamma[{\mathbf r}(t)]} \;=\; 
\left\langle   
	e^{i{\mathcal S}_{\rm I}(A; {\mathbf r}) }
\right\rangle 
\end{equation}
with:
\begin{equation}
\langle \ldots \rangle \;\equiv\; 
\frac{\int {\mathcal D}A \;\ldots\; 
	e^{i {\mathcal S}_0(A)}}{\int {\mathcal D}A \; e^{i {\mathcal S}_0(A)}} \;.
\end{equation}

$\Gamma$  may be expanded in powers of ${\mathcal S}_{\rm int}$, producing
a series of terms, namely, $\Gamma = \Gamma_1 + \Gamma_2 +
\ldots$. 
Let us consider the first two of them:
\subsection{First order contribution}
We see that the first-order contribution, $\Gamma_1$, contains two terms,
\begin{equation}
\Gamma_1 \,=\, \Gamma_1^{(a)} + \Gamma_1^{(g)} \;,
\end{equation}
with
\begin{equation}
	\Gamma_1^{(a,g)} \;=\; \langle {\mathcal S}_{\rm I}^{(a,g)}(A; {\mathbf r})
	\rangle \;.
\end{equation}
It is clear that $\Gamma_1^{(g)}$ corresponds to a vacuum energy
contribution, which is a well-known result for graphene. 

 We now consider $\Gamma_1^{(a)}$. We see that:
\begin{equation}
	\Gamma_1^{(a)} \;=\; \frac{i}{2} e^2 \, \int dt \int dt' \;  
	\Delta_\Omega(t-t') \; 
	\langle {\mathbf E}(t,{\mathbf r}(t)) \cdot {\mathbf E}(t',{\mathbf
	r}(t'))  \rangle \;,
\end{equation}
which involves the free electric field correlation function. This, may be
expressed as follows:
\begin{align}\label{eq:ee}
	\langle  E_i (t,{\mathbf x})  E_j (t',{\mathbf x'})  \rangle  &=\;
	i \, \delta_{ij} \, \delta(t-t') \, \delta({\mathbf x}-{\mathbf
	x'}) \nonumber\\
	&+\,i \, \int \frac{d^4k}{(2\pi)^4} e^{-i k \cdot ({\mathbf x} -{\mathbf x'})}\, 
\; \frac{{\mathbf k}^2 \delta_{ij} - k_i k_j }{k_0^2
	- {\mathbf k}^2  + i \epsilon} \;.
\end{align}

The first line in the electric-field correlation, when introduced in the effective action,  induces a divergent energy
shift. If ${\mathit l}$ is
a length characterizing the size of the atom, the energy shift reads  $E_{\rm div} = \frac{ 3 e^2}{ 4 m \Omega {\mathit l}^3}$. 

Aside from this contribution, the second, finite one, may be conveniently
studied in Fourier space, where it can be related to our previous results
on the scalar model~\cite{Belen2019}.
Introducing:
\begin{equation}\label{eq:f}
f({\mathbf p},\nu) \,=\,
\int_{-\infty}^{+\infty} dt \,  e^{-i {\mathbf p}\cdot {\mathbf r}(t)} \;
e^{i\nu t} \;,
\end{equation}
the result for the finite part is,
\begin{equation}
	\Gamma_1^{(a)} \;=\; \frac{1}{2}\,\int \frac{d\nu}{2\pi} \, 
\int \frac{d^3p}{(2\pi)^3}\, f(-{\mathbf p}, -\nu)
f( {\mathbf p}, \nu) \;
\Pi(\nu,{\mathbf p}) \;,
\end{equation}
where
\begin{equation}\label{eq:defpi1p}
\Pi(\nu,{\mathbf p}) \;\equiv\; -  \frac{ 2 i e^2}{m} \, {\mathbf p}^2 \, 
\int \frac{dp^0}{2\pi} \; \frac{1}{(p^0 - \nu)^2 - \Omega^2 + i
\epsilon} \,\frac{1}{(p^0)^2 - {\mathbf p}^2 + i \epsilon} \;.
\end{equation}

Therefore, 
\begin{equation}
{\rm Im} [ \Gamma_1^{(a)}]\;=\; \frac{1}{2} \,
	\int \frac{d\nu}{2\pi} \int \frac{d^3p}{(2\pi)^3} \;
	\big| f({\mathbf p}, \nu) \big|^2 \;
	{\rm Im}\big[\Pi(\nu, |{\mathbf p}|) \big] \;,
\end{equation}
where: 
\begin{equation}\label{eq:impi}
	{\rm Im}\big[\Pi(\nu, |{\mathbf p}|) \big] \;=\; \frac{\pi e^2
	|{\mathbf p}| }{ m \Omega} \,
	\big[ \delta( \nu - |{\mathbf p}| - \Omega) +  \delta( \nu + |{\mathbf p}| + \Omega)
	\big] \;.
\end{equation}

From (\ref{eq:impi}), we note that there is a threshold in the frequency
$\nu$, below which there is no photon emission. That is a threshold for a
Fourier component of frequency $\nu$ in $f({\mathbf p},\nu)$, {\em which is not
necessarily the same as the frequency of motion}, unless one expands the
expressions in powers of the oscillation amplitudes.  Indeed, recalling the
definition of $f$, we see that for a small, bounded motion ${\mathbf r}(t)
= {\mathbf r}_0 + {\mathbf y}(t)$, where ${\mathbf r}_0$ is the average
position and ${\mathbf y}(t)$ the departure, a series expansion yields, up
to the lowest non trivial order:
\begin{equation}
	f({\mathbf p},\nu) \;=\; e^{-i {\mathbf p}\cdot {\mathbf r}_0} \left[ 2 \pi \,
	\delta(\nu) \,-\,i {\mathbf p} \cdot \tilde{\mathbf y}(\nu) \right]
	\;,  
\end{equation}
where $\tilde{\mathbf y}(\nu)$ is the Fourier transform of ${\mathbf
y}(t)$. Since the time average of ${\mathbf y}(t)$ vanishes,
$\tilde{\mathbf y}(0)=0$. Upon insertion of this expansion for $f$ into the
imaginary part, we can integrate the spatial components of the momentum, to
get, at the second order in the departure, the spectral form:
\begin{equation}\label{eq:imord1}
{\rm Im} [ \Gamma_1^{(a)}]\;=\; \frac{1}{2} \,
	\int \frac{d\nu}{2\pi} \, m^{ij}(\nu) \; \tilde{y}^i(-\nu)
	\tilde{y}^j(\nu) \;,
\end{equation}
where:
\begin{equation}\label{res:order11}
	m^{ij}(\nu) \;=\; \frac{e^2}{6 \pi m \Omega} \, \delta^{ij} \;
	\theta(|\nu| - \Omega) \, (|\nu| - \Omega)^5  \;.
\end{equation}
By dimensional reasons, the spectrum has a different power law that the
scalar counterpart of this model \cite{Belen2019}, as well as different coefficient and
factor of $2$ (due to the polarizations of the EM field).

In particular, for a small linear oscillatory motion with frequency $\nu_0$
and amplitude ${\mathbf s}$,
${\mathbf y}(t) = {\mathbf s} \, {\rm cos}(\nu_0 t)$, we obtain a constant
${\rm Im}[\Gamma_1^{(a)}]$ per unit time (vacuum decay rate):
\begin{equation}\label{eq:imgpert}
	\frac{{\rm Im}[\Gamma_1^{(a)}]}{T}\;=\; {\mathbf s}^2 \,
	\frac{e^2}{24 \pi m \Omega} \,
	\theta(|\nu_0| - \Omega) \, (|\nu_0| - \Omega)^5 \;,
\end{equation}
where $T$ is the total time.

So far, this corresponds to the second order in the oscillation amplitude,
always within the $\Gamma_1^{(a)}$ contribution. When considering the same
oscillatory motion without expanding in powers of the amplitude, we note
that $f$ contains not just the oscillation frequency $\nu_0$ but also its 
harmonics. Indeed, using the Jacobi-Anger expansion,
\begin{equation}
	e^{-i {\mathbf p} \cdot {\mathbf r}(t)} \,=\, e^{-i {\mathbf p}
	\cdot {\mathbf s}\, {\rm cos}(\nu_0 t) } \,=\,
	\sum_{n=-\infty}^{+\infty} \, (-i)^n \, J_n ({\mathbf p} \cdot {\mathbf s}) \,
	e^{- i n \nu_0 t} \;,
\end{equation}
(where $J_n$ denotes a Bessel function of the first kind) we see that all
the  multiples of $\nu_0$ are going to be present in $f$, what
means that the threshold $\Omega$ may be surpassed with a lower
oscillation amplitude (if its amplitude is increased):
\begin{equation}
	f({\mathbf p}, \nu) \,=\, 2 \pi \, \sum_{n=-\infty}^{+\infty} \,
	(-i)^n \, J_n ({\mathbf p} \cdot {\mathbf s}) \, \delta(\nu - n
	\nu_0) \;.
\end{equation}
It is interesting to note that, if one considers  oscillations involving more than one
direction, like for example when performing a circular motion, then the
phenomenon above involves sums of the respective frequencies, since the
expansion above produces a series for each factor. 

For the single linear oscillation, we find:
\begin{equation}\label{eq:imgnp}
\frac{{\rm Im}[\Gamma_1^{(a)}]}{T}\;=\; \frac{e^2}{\pi m \Omega} \,
	\sum_{n=1}^\infty \, \left\{ \theta(n \nu_0 - \Omega) \, (n \nu_0 - \Omega)^3 \;
	\int_0^1 du \, J_n[ (n \nu_0 - \Omega) |{\mathbf s}|
	\,  u ] \right\} \;.
\end{equation}
A few comments are in order; firstly, the term quadratic in the oscillation
amplitude is recovered if one assumes that $\nu_0$ is above the threshold,
and one expands in powers of the amplitude each Bessel function.  The $n=1$
terms produces the term quadratic in the departure. 

On the other hand, if the oscillation frequency is lower than the
threshold, one gets a contribution starting from $n > 1$.  The relevance of the harmonics of the fundamental frequency in 
the DCE  has been previously pointed out in Ref.\cite{Giraldo}, for the case of semitransparent mirrors.

\subsection{Second order}\label{ssec:second}
The second-order term is found to be given by:
\begin{equation}
	\Gamma_2\;=\; \frac{i}{2} \, \langle \big( {\mathcal S}_I - \langle
	{\mathcal S}_I \rangle \big)^2 \rangle \;=\;
	\Gamma_2^{(aa)} \,+\,\Gamma_2^{(gg)} \,+\, \Gamma_2^{(ag)} \;,
\end{equation}
of which, in a self-explanatory notation, only the last one involves
correlations between the atom and the graphene plate (the first one deals
with the atom in free space, and the second one produces a contribution to
the graphene self-energy):
\begin{equation}
	\Gamma_2^{(ag)}\;=\; i \, \langle 
	\big( {\mathcal S}_I^{(a)} - \langle {\mathcal S}_I^{(a)}\rangle \big)
	\big( {\mathcal S}_I^{(g)} - \langle {\mathcal S}_I^{(g)}\rangle \big)
	\rangle \;.
\end{equation}
Introducing the explicit form of ${\mathcal S}_I^{(a)}$ and ${\mathcal
S}_I^{(g)}$, keeping the latter in terms of a yet unspecified vacuum
polarization tensor, after some algebra we get:
\begin{equation}
	\Gamma_2^{(ag)} \;=\; \frac{1}{2} \, \int
	\frac{d^2{\mathbf k_\parallel}}{(2\pi)^2}  \int\frac{dk_3}{2\pi}
\int\frac{dp_3}{2\pi} \int\frac{d\nu}{2\pi} \;
f(-{\mathbf k_\parallel},k^3, -\nu) 
f({\mathbf k_\parallel},p^3,\nu) 
	\; B(\nu,{\mathbf k_\parallel},k^3,p^3)
\end{equation}
where
\begin{equation}\label{eq:defb}
B(\nu,{\mathbf k_\parallel},k^3,p^3) \;=\; e^2 \; 
	\int\frac{dk_0}{2\pi} \; \frac{\widetilde{\Delta}_\Omega(k_0 + \nu) \, 
	\big[\widetilde{\Pi}_{ii}(k_\shortparallel) k_0^2 + 
	\widetilde{\Pi}_{00}(k_\shortparallel) ({\mathbf
	k_\shortparallel}^2 - k_3 p_3)  
	- 2 k_0 k_i \widetilde{\Pi}_{0i}(k_\shortparallel)\big]}{ \big(k_0^2 -
	{\mathbf k_\shortparallel}^2-k_3^2\big) \big(k_0^2 - {\mathbf
	k_\parallel}^2-p_3^2\big)}\,. 
\end{equation}
In terms of the functions $g_t$ and $g_l$,
\begin{equation}\label{eq:defb1}
B(\nu,{\mathbf k_\shortparallel},k^3,p^3) \;=\; - e^2 \; 
	\int\frac{dk_0}{2\pi} \; \frac{\widetilde{\Delta}_\Omega(k_0 + \nu) \, 
	\big[ k_0^2 g_t(k_\shortparallel) + k_\shortparallel^2 g_l(k_\shortparallel) 
	- \frac{{\mathbf k_\shortparallel}^2}{k_\shortparallel^2} k_3 p_3
	g_l(k_\shortparallel)\big]}{ \big(k_0^2 - {\mathbf
	k_\shortparallel}^2-k_3^2\big) \big(k_0^2 - {\mathbf
	k_\parallel}^2-p_3^2\big)}\,. 
\end{equation}
For the particular case of graphene, those functions  may be conveniently
represented by means of an  integral representation:
\begin{align}
	g_t(k_\shortparallel) &=\; 2 \alpha_N \, (k_0^2 - v_F^2 {\mathbf
	k_\shortparallel}^2) \, \int \frac{dq_3}{2\pi} \, \frac{1}{k_0^2 -
	v_F^2 {\mathbf k_\shortparallel}^2 - q_3^2} \nonumber\\
	g_l(k_\shortparallel) &=\; 2 \alpha_N \, (k_0^2 - {\mathbf
	k_\shortparallel}^2) \, \int \frac{dq_3}{2\pi} \, \frac{1}{k_0^2 -
	v_F^2 {\mathbf k_\shortparallel}^2 - q_3^2} \;,
\end{align}
which, when used into (\ref{eq:defb1}) leads to:
\begin{align}\label{eq:defb2}
B(\nu,{\mathbf k_\shortparallel},k^3,p^3) =\; - 2 \, i \,  \frac{e^2}{m} \, \alpha_N \, 
	\int\frac{dq_3}{2\pi} \int\frac{dk_0}{2\pi} \; \big[ k_0^2
	(k_0^2 - v_F^2 {\mathbf k_\shortparallel}^2) +
	(k_\shortparallel^2)^2 - {\mathbf k_\shortparallel}^2 k_3 p_3
	\big] \nonumber\\
	\times  \frac{1}{\big((k_0 + \nu)^2 - \Omega^2 + i \epsilon\big)
	\big(k_0^2 - {\mathbf k_\shortparallel}^2-k_3^2 + i \epsilon \big) \big(k_0^2 -
	{\mathbf k_\shortparallel}^2-p_3^2 + i \epsilon\big) \big(k_0^2 -
	v_F^2 {\mathbf k_\shortparallel}^2-q_3^2 + i \epsilon \big)}\;, 
\end{align}
where we have written explicitly the $i \epsilon$ terms in the
denominators.
The next step is to integrate $k_0$, which appears in a similar fashion as
the momentum to integrate in a loop integral in a quantum field theory
system, although in this case in $0+1$ dimensions. This can be done in several way, for
example by the method of residues. 
After that integration, the imaginary part of $B$ may be written as
follows:
\begin{align}\label{eq:imb}
	{\rm Im}\Big[B(\nu,{\mathbf k_\shortparallel},k^3,p^3)\Big] \;=\;
	\frac{\pi e^2\alpha_N}{m \Omega}  
	\int\frac{dq_3}{2\pi} \;\Big\{ &
	\frac{(1-v_F^2) |{\mathbf k_\shortparallel}|^4 
	+ [(2 -v_F^2) p_3^2 - p_3 k_3] |{\mathbf k_\shortparallel}|^2 + 2
	p_3^4}{(p_3^2 - k_3^2) \big[(1 -v_F^2) |{\mathbf k_\shortparallel}|^2 +
	p_3^2 - q_3^2\big] p}  \nonumber\\
	\times \big[ \delta(\nu - \Omega- p) + \delta(\nu +
	\Omega+ p)\big] &+ \frac{(1-v_F^2) |{\mathbf k_\shortparallel}|^4 
	+ [(2 -v_F^2) k_3^2 - k_3 p_3] |{\mathbf k_\shortparallel}|^2 + 2
	k_3^4}{(k_3^2 - p_3^2) \big[(1 -v_F^2) |{\mathbf k_\shortparallel}|^2 +
	k_3^2 - q_3^2\big] k}  \nonumber\\
	\times 
	\big[ \delta(\nu - \Omega- k) + \delta(\nu +
	\Omega+ k)\big] 
	&+\frac{(1-v_F^2)^2 |{\mathbf k_\shortparallel}|^4 
	+ [(-2 + 3 v_F^2) q_3^2 - k_3 p_3] |{\mathbf k_\shortparallel}|^2 + 2
	q_3^4}{\big[(1 -v_F^2) |{\mathbf k_\shortparallel}|^2 +
	k_3^2 - q_3^2\big] \big[(1 -v_F^2) |{\mathbf k_\shortparallel}|^2 +
	p_3^2 - q_3^2\big] q}  \nonumber\\
	\times \big[ \delta(\nu - \Omega - q) + \delta(\nu +
	\Omega+ q)\big] & \Big\} \;
\end{align}
where we have introduced the shorthand notation: \mbox{$p \equiv \sqrt{{\mathbf
k_\shortparallel}^2 + p_3^2}$}, \mbox{$k \equiv \sqrt{{\mathbf k_\shortparallel}^2 +
k_3^2}$}, and \mbox{$q \equiv \sqrt{v_F^2 {\mathbf k_\shortparallel}^2 +
q_3^2}$}.

Note that the third term inside the integral can be integrated, by taking
advantage of the $\delta$-function involving $q_3$. 

Regarding the first and second terms, we see that they differ just in the
interchange $k^3 \leftrightarrow p^3$, so we discuss just the first one: 
the sum of the two $\delta$-functions may be replaced by the product:
$\theta(|\nu| - \Omega) \; \delta(|\nu|- \Omega - p)$.
Therefore, using the previous condition into the integral for the (only)
factor involving $q_3$,
\begin{align}
	\int \frac{dq_3}{2\pi} 	\frac{1}{(1 -v_F^2) |{\mathbf k_\shortparallel}|^2 +
	p_3^2 - q_3^2} \;=\;
	\int \frac{dq_3}{2\pi} \, \frac{1}{-v_F^2 |{\mathbf k_\shortparallel}|^2 +
	(|\nu| - \Omega)^2  - q_3^2} \nonumber\\
	\;=\; - \, \theta\big(v_F |{\mathbf k_\shortparallel}|-
	(|\nu|-\Omega)\big) \; \frac{1}{2  \sqrt{v_F^2 |{\mathbf
	k_\shortparallel}|^2 - (|\nu|-\Omega)^2}} \;,
\end{align}
where the last equality follows from the use of a principal value
prescription for the integral. Note however that
the condition $p=|\nu|- \Omega$ forced by the $\delta$-function implies
that  $v_F
|{\mathbf k_\shortparallel}| < |\nu|-\Omega$ for $v_F < 1$. Therefore,
the first and second terms in Eq.\eqref{eq:imb} do vanish.

Using this result, we see that the full imaginary part
may be written as follows:
\begin{align}\label{eq:imbfin}
{\rm Im}\Big[B(\nu,{\mathbf k_\shortparallel},k^3,p^3)\Big] &=\;
		 \frac{\pi e^2\alpha_N}{m \Omega} \theta(|\nu| - \Omega) 
			 \; \theta(|\nu| - \Omega - v_F |{\mathbf
		k_\shortparallel}|) \nonumber\\
	&\times \;\frac{[|{\mathbf k_\shortparallel}|^4 - [ (2 +  v_F^2) (|\nu|-
\Omega)^2 + k_3 p_3 ]|{\mathbf k_\shortparallel}|^2 + 2 (|\nu|-
	\Omega)^4 ]}{\big[|{\mathbf k_\shortparallel}|^2 +
k_3^2 - (|\nu|- \Omega)^2 \big]  \big[|{\mathbf k_\shortparallel}|^2 +
p_3^2 - (|\nu|- \Omega)^2 \big] (|\nu|- \Omega)} 
	 \;.
\end{align}
In what follows we will apply  this result to the analysis of the dissipative effects on the motion of the atom induced by the presence of a graphene sheet.

\section{Results}\label{sec:results}
\subsection{Quantum friction}
The first example that we consider corresponds to an atom
moving with constant velocity ${\mathbf u}$, parallel the plane.  A non vanishing imaginary part of the effective action would indicate the presence 
of quantum friction produced by the graphene.

The trajectory of the atom is given bay ${\mathbf r}(t) = (0, u t , a)$.  
From Eq.\eqref{eq:f} we obtain
\begin{equation}
	f({\mathbf k},\nu) \,=\,2 \pi \,e^{-i k^3 a } \, \delta(\nu - k^1 u) \;.
\end{equation}
Therefore, denoting by $T$ the total extent of the time interval, we see
that the imaginary part of the effective action per unit time becomes:
\begin{equation}\label{eq:friction1}
	{\rm Im}\Big[\frac{\Gamma_2^{(ag)}}{T}\Big] \;=\;\frac{1}{2} \,
	\int \frac{dk^1}{2\pi} \frac{dk^2}{2\pi} \frac{dk^3}{2\pi} \frac{dp^3}{2\pi}
	e^{-i (p^3 + k^3) a} \; {\rm Im}\Big[B(k^1 u,{\mathbf k_\shortparallel},k^3,p^3)\Big] \;.
\end{equation}
We see that, due to the presence of the threshold in the imaginary part of
$B$, that {\em there will not be friction for $u < v_F$}. 

Note, however, that $u > v_F$, there will be a non-vanishing imaginary
part. The physical reason is that for that velocity, the frequency involved
($k^1 u$) can excite physical Dirac fermions on the graphene sheet. 
Note also  that, as the velocity of the atom is $u<1$,  the integrals over $p^3$ and $k^3$ are well-defined, since
the denominator in Eq.\eqref{eq:imbfin} never vanishes. 

After integrating those variables, we get:
\begin{align}\label{eq:friction2}
{\rm Im}\Big[\frac{\Gamma_2^{(ag)}}{T}\Big] &=\;\frac{\pi e^2\alpha_N}{8 m \Omega} 
	\int \frac{dk^1}{2\pi} \frac{dk^2}{2\pi} 
		\; \theta[|k^1 u | - \Omega - v_F |{\mathbf
		k_\shortparallel}|] \, e^{- 2 a \sqrt{|{\mathbf
		k_\shortparallel}|^2 - (|k^1 u|- \Omega)^2}} \nonumber\\
	&\times \;\frac{ 2 |{\mathbf k_\shortparallel}|^4 - (3 +  v_F^2)
	(|k^1 u|- \Omega)^2 |{\mathbf k_\shortparallel}|^2 + 2 (|k^1 u|-
	\Omega)^4 }{\big[|{\mathbf k_\shortparallel}|^2 - (|k^1 u|- \Omega)^2 \big] (|k^1 u |- \Omega)} 
	 \;.
\end{align}
One can verify that, under the assumptions mentioned before ($|{\mathbf
k_\shortparallel}| >  |k^1 u | - \Omega >  v_F |{\mathbf
k_\shortparallel}|$), the integrand is positive. 
\begin{figure}
\includegraphics[width=10cm]{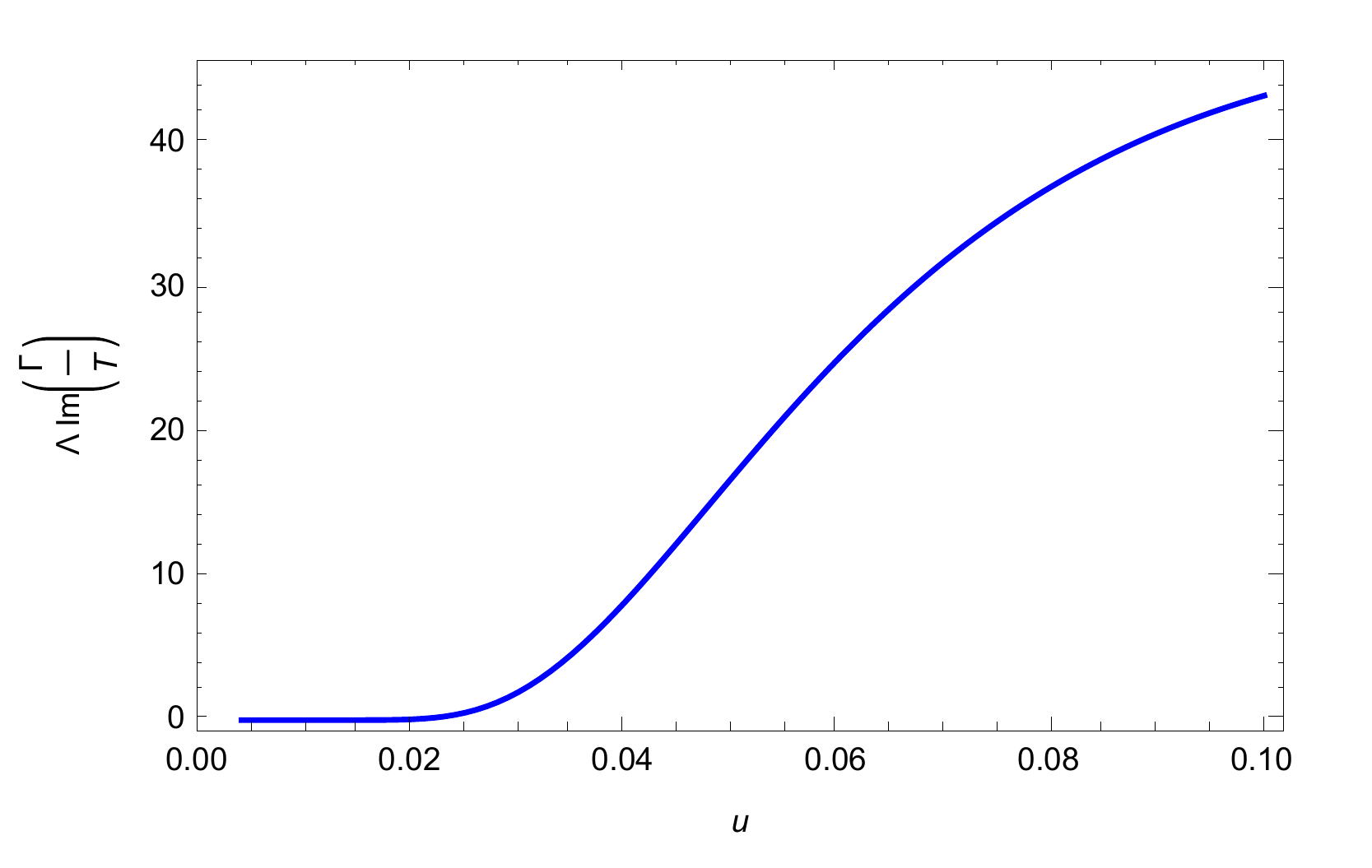}
\caption{Imaginary part of the effective action as a function of the atom
	velocity $u$ (dimensionless in natural units). There is not
	dissipative effect until $u > v_F = 0.003$, and QF increases as the
parallel speed of the atom grows. We have set $\Lambda = 8a^3 m\Omega/(\pi
e^2 \alpha_N)$ and $\Gamma_2^{(a,g)} = \Gamma$.}  
\label{fig1} 
\end{figure}
In Fig. 1 we plot the imaginary part of the effective action as a function
of the velocity of the atom. We see that, after the threshold and an
initial plateau, quantum friction increases with the velocity. The results
are qualitatively similar to those obtained for the quantum friction
between graphene sheets in Ref.\cite{Belen2017}. In that reference, it was
also shown how to obtain the frictional force from the imaginary part of the
in-out effective action (\ref{eq:friction1}). 
Besides, the argument for the existence of a threshold presented in
Ref.\cite{Belen2017}, for the case of two sliding graphene sheets, can be 
translated here in a rather direct way:  consider the momentum and energy
balance during a small time interval $\delta t$, assuming that both the frictional force and the
dissipated energy are driven by pair creation. 
The only relevant component of the total momentum ${\mathbf P}$ of the pair
for the (momentum) balance is the one along the direction of the velocity
$v$. Relating that component of ${\mathbf P}$ to the frictional
force, we see that: 
\begin{equation}\label{eq:bmom}
F_\text{fr} \, \delta t \; = \;  P_x \;.
\end{equation}
On the other hand, the energy balance reads
\begin{equation}\label{eq:bener}
F_\text{fr} \, v \, \delta t \; = \; {\mathcal E}
\end{equation}
where ${\mathcal E}$ is the energy of the pair. But, since the fermions are both
on-shell, we have 
\begin{equation}\label{eq:disper}
{\mathcal E} \geq v_F |P_x| 
\end{equation}
(the equal sign corresponds to a pair with momentum along the direction of
$v$).  Dividing Eq.(\ref{eq:bener}) and Eq.(\ref{eq:bmom}), and taking into account
Eq.(\ref{eq:disper}), we see that a necessary condition for friction to
happen is $v \; \geq \; v_F \;$.

The situation considered here is more interesting for an eventual
experimental observation of the effect, because frictional effects only
exist if the speed of the sliding motion is larger than the Fermi velocity
of the charge carriers in graphene, a condition that is  more easily
achieved for an atom than for a whole sheet (for example for the
implementation of the experimental proposal in \cite{qute21}).

\subsection{Small departures normal to the graphene plane}\label{sub:perp}

In Section IV.A we computed the vacuum persistence amplitude for an atom oscillating in vacuum. The result  to lowest order in the amplitude is
given in Eq.\eqref{eq:imgpert}, and without expanding in the amplitude
in Eq.\eqref{eq:imgnp}. 

Here we will consider the corrections to the vacuum persistence amplitude
induced by the presence of a graphene sheet. To this end, we will evaluate the imaginary part of $\Gamma_2^{(ag)}$ in a
situation where the atom undergoes a bounded motion, always in a direction
normal to the graphene. Although the expression for the imaginary part is
not perturbative in the amplitude of the atom's motion, we will find here
the result to the lowest order in that amplitude. 

We assume ${\mathbf r}(t) = {\mathbf r}_0 + {\mathbf y}(t)$, where the mean
position of the atom is ${\mathbf r}_0 = (0,0,a)$, with $a > 0$. The
departure, on the other hand, is ${\mathbf y}(t) = (0,0, y_\perp (t))$. 
We just need to insert the corresponding $f$ for this situation in the
general expression; to the order we want to work, it is sufficient to use:
\begin{equation}
	f({\mathbf k},\nu) \;=\; \,e^{-i k^3 a} \, [ 2 \pi \delta(\nu) \,-\,
	i k^3 y_\perp(t) \,+\, \ldots] \;.
\end{equation}
The resulting imaginary part may be arranged, in a similar fashion as is
(\ref{eq:imord1}), as follows:
\begin{equation}
{\rm Im} [ \Gamma_2^{(ag)}]\;=\; \frac{1}{2} \,
	\int \frac{d\nu}{2\pi} \, m_\perp(\nu) \; |\tilde{y}_\perp(\nu)|^2 \;,
\end{equation}
where,
\begin{equation}\label{eq:mperp}
m_\perp(\nu)  \;=\; - \, \int \frac{d^2{\mathbf k_\parallel}}{(2\pi)^2}  \int\frac{dk^3}{2\pi}
	\int\frac{dp^3}{2\pi} \, k^3 \, p^3\, e^{-i (k^3 + p^3) a} \; 
	\; {\rm Im}\big[B(\nu,{\mathbf k_\parallel},k^3,p^3) \big]\;.
\end{equation}

Unlike the case of quantum friction described in the previous section, in the calculation of $m_\perp^{(A)}(\nu)$ the evaluation of 
the $k^3$ and $p^3$ integrals should be handled with care, using a principal value prescription. We include some details of the calculation in the Appendix. 
The result of performing the integrals  allows us to write 
\begin{equation}\label{eq:mperpfin}
	m_\perp(\nu) \;=\; \frac{ e^2 \alpha_N}{8 m \Omega} \,
	\theta(|\nu| - \Omega) \, \big(|\nu| - \Omega\big)^5 \; 
	\varphi \big(|\nu| a - \Omega a \big) \;,
\end{equation}
where $\varphi (x)$  is a dimensionless function (of a dimensionless
variable), given by:
\begin{align}\label{eq:varphiint}
	\varphi (x) &=\;  \int_0^{v_f^{-2}} d\rho \big(
		1 - \frac{3 + v_F^2}{2}\rho + \rho^2  \big) \big [\cos( 2 x
	\sqrt{1-\rho}) \theta(1-\rho)\nonumber\\
	&+\;  e^{- 2 x \sqrt{\rho -1}}\theta(\rho-1) \big]\;.
\end{align}
These integrals can be computed analytically in terms of elementary functions. To do this it is useful to perform
the change of variables $y=\sqrt{|\rho -1|}$. 
The explicit expression reads
\begin{equation}\label{eq:varphian}
\varphi(x)=\varphi_1(x)+\varphi_2(x)\, ,
\end{equation}
where
\begin{align}
\varphi_1(x)&=\frac{1}{8 x^6}\Big[-30+
\left(v_F^2-1\right) \left(2 x^2+3\right) x^2+2 x \left(-\left(3 v_F^2+17\right) x^2+4 x^4+30\right) \sin (2 x)\nonumber\\
&+\left(4
   \left(v_F^2+4\right) x^4-3 \left(v_F^2+19\right) x^2+30\right) \cos (2 x)\Big]\, ,
\end{align}
and
\begin{align}
\varphi_2(x)&=\frac{e^{-\frac{2 \sqrt{1-v_F^2} x}{v_F}}}{8\, v_F^5 \,x^6}
\Big[v_F^7 x^2 \left(3-4 x^2\right)+v_F^5 \left(-10 x^4+57 x^2-30\right)+v_F^3 \left(34 x^4-60 x^2\right)\nonumber\\
&-8 \sqrt{1-v_F^2}\, x^5+4
   \sqrt{1-v_F^2} \,v_F^2 x^3 \left(3 x^2-10\right)+6 \sqrt{1-v_F^2}\, v_F^6 \,x^3
   \nonumber\\
   &-2 \sqrt{1-v_F^2}\, v_F^4\, x \left(x^2-6\right) \left(2
   x^2-5\right)-20\,  v_F\, x^4\Big]\nonumber\\
   &+\frac{1}{8 x^6}\left(30-\left(v_F^2-1\right) x^2 \left(2 x^2+3\right)\right)\, .
\end{align}
Some properties of the function $\varphi(x)$ can be derived from its analytic expression. For instance, in the limit
$x\to 0$ it tends to the large finite value $\varphi(0)\simeq 1/(3 \, v_F^6)$, while in the large $x$ limit $\varphi(x)\simeq x^{-1}\sin(2 x)$.

In Fig. 2 we plot the function $\varphi$ as a function of $x =  a
\big(|\nu| - \Omega\big)$. As anticipated, it reaches large amplitudes at very small 
values of $x$, but then approaches zero, oscillating with a small amplitude,
for large $x$. The behaviour of $m_\perp(\nu)$ at extremely small distances $a$ 
between the atom and the graphene sheet is determined by that of $x^5
\varphi(x)$,  as can be seen in Fig. 3. 
\begin{figure}
\includegraphics[width=9cm]{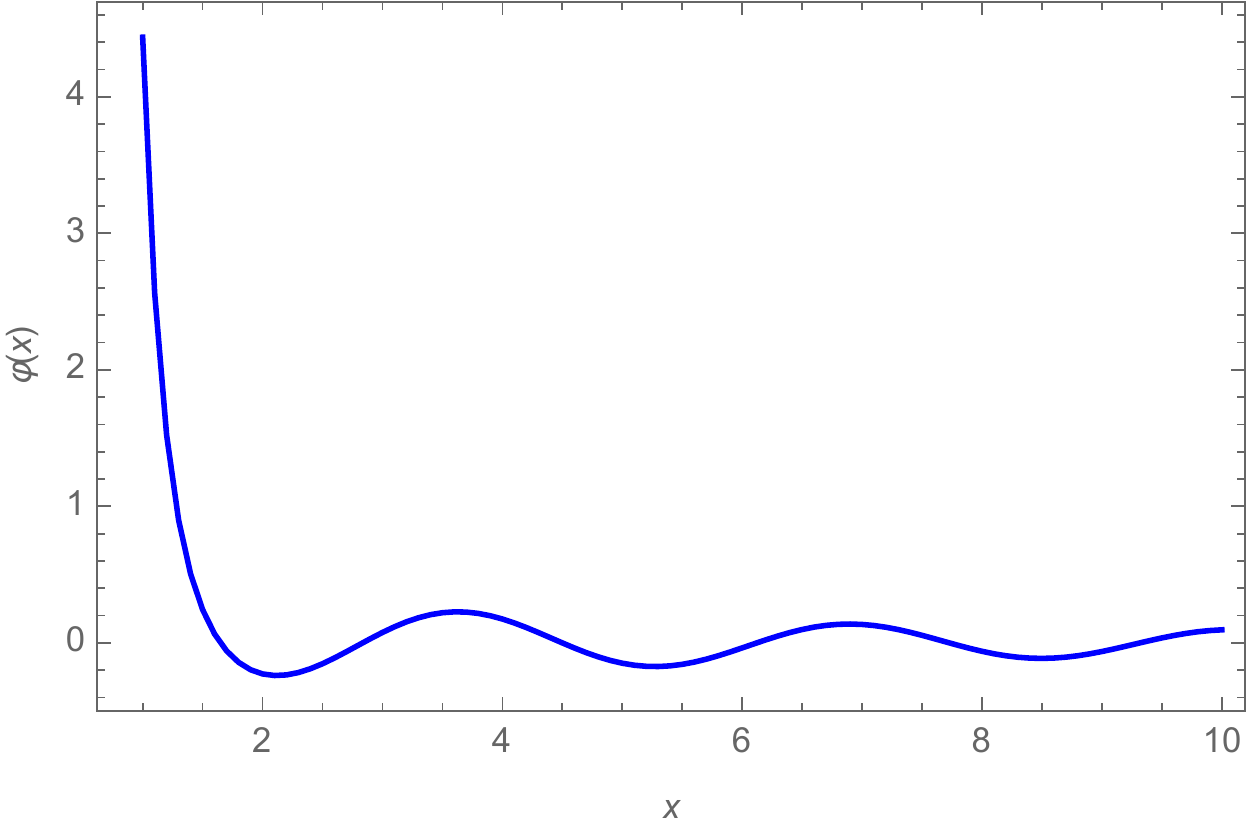}
\caption{Dimensionless function $\varphi$ as a function of $x =  a \big(|\nu| - \Omega\big)$ (also dimensionless). }  \label{fig2}
\end{figure}
\begin{figure}
\includegraphics[width=9cm]{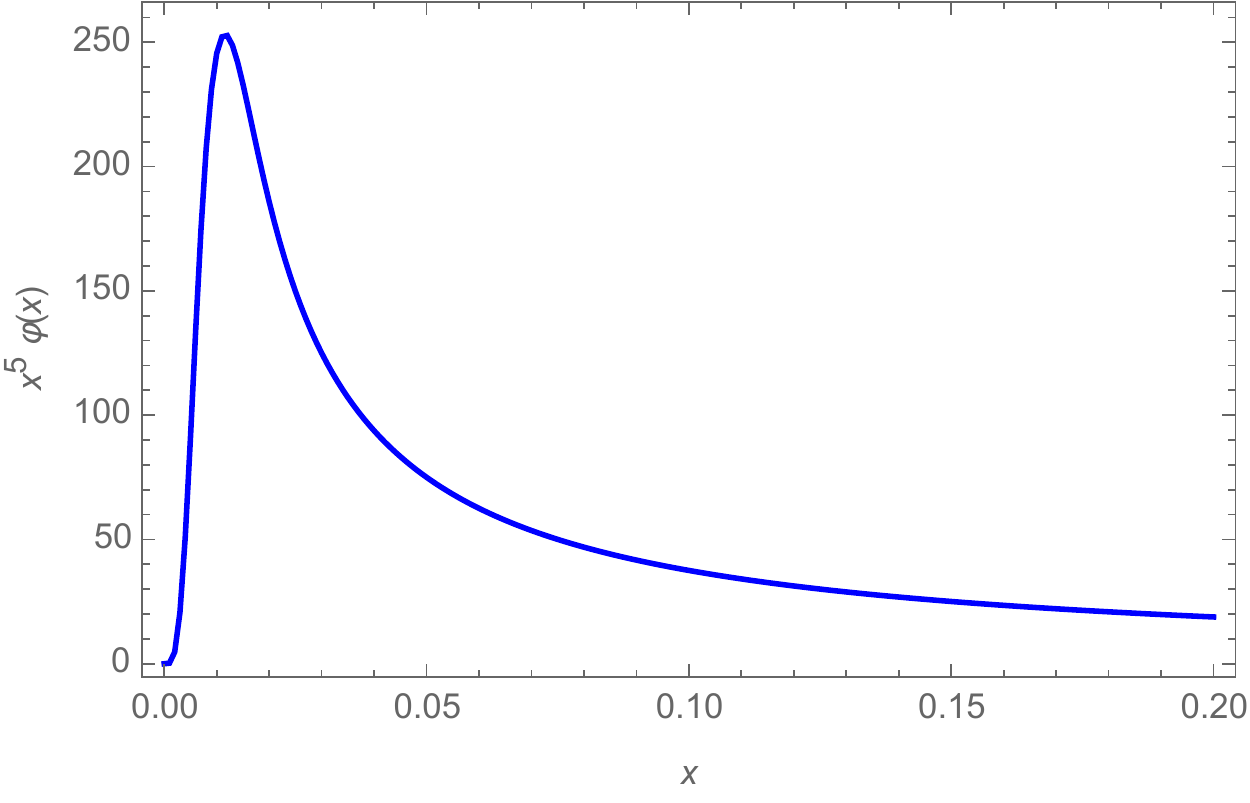}
\caption{Small $x$ behaviour of the product $x^5 \varphi(x)$. 
}  \label{fig3}
\end{figure}

\subsection{Small departures parallel to the graphene plane}\label{sub:paral}

We will now consider oscillations of the atom that are parallel to the
graphene sheet. 
The trajectory of the particle is now given by  ${\mathbf r}(t) = {\mathbf r}_0 + {\mathbf y}(t)$, where, as before,  ${\mathbf r}_0 = (0,0,a)$, with $a > 0$. The
departure from the mean position is 
${\mathbf y}(t) = ({\mathbf y_\shortparallel}(t),0)$ and the corresponding
$f$ function reads:
\begin{equation}
	f({\mathbf k},\nu) \;=\; \,e^{-i k^3 a} \, [ 2 \pi \delta(\nu) \,-\,
	i {\mathbf k_\shortparallel}\cdot \tilde{\mathbf 
	y_\shortparallel}(\nu) \,+\, \ldots] \;.
\end{equation}
The imaginary part of the effective action  reads
\begin{equation}
{\rm Im} [ \Gamma_2^{(ag)}]\;=\; \frac{1}{2} \,
	\int \frac{d\nu}{2\pi} \, m_\parallel(\nu) \; |\tilde{\mathbf y_\shortparallel}(\nu)|^2 \;,
\end{equation}
where
\begin{equation}
m_\parallel(\nu)  \;=\; \frac{1}{2} \, \int \frac{d^2{\mathbf k_\parallel}}{(2\pi)^2}  \int\frac{dk^3}{2\pi}
	\int\frac{dp^3}{2\pi} \, |{\mathbf k_\shortparallel}|^2 e^{-i (k^3 + p^3) a} \; 
	\; {\rm Im}\big[B(\nu,{\mathbf k_\parallel},k^3,p^3) \big]\;.
\end{equation}
The formal expression for $m_\parallel(\nu)$ is rather similar to that of $m_\perp (\nu)$,
Eq.\eqref{eq:mperp}, with the replacement $k^3p^3\to -1/2 |{\mathbf
k_\shortparallel}|^2$. Its evaluation proceeds along rather similar steps,
and we therefore  omit the details. The final result reads:
\begin{equation}\label{eq:mparalfin}
	m_\parallel(\nu) \;=\; -\frac{ e^2 \alpha_N}{16 m \Omega} \,
	\theta(|\nu| - \Omega) \, \big(|\nu| - \Omega\big)^5 \; 
	\phi \big(|\nu| a - \Omega a \big) \;,
\end{equation}
where 
\begin{eqnarray}\label{eq:phiint0}
\phi(x)&=& PV \int_0^{v_F^{-2}}d\rho\, \frac{\rho}{1-\rho} \big(\rho^2-\frac{3+v_F^2}{2}\rho + 1\big)\nonumber\\
&\times &\big[\cos(2  x\sqrt{1-\rho})\theta(1-\rho)+ e^{-2  x \sqrt{\rho-1}}\theta(\rho - 1)\big]\, .
\end{eqnarray}
The integral that defines the function $\phi(x)$ has a singularity at $\rho=1$. However, it is well defined with the principal value prescription.

We highlight the relation between this result and the function
$\varphi(x)$ in Eq.\eqref{eq:varphiint}. Note that the difference  is the extra factor
$\rho/(1-\rho)$ in the integrand of Eq.\eqref{eq:phiint0}. Therefore we write
\begin{eqnarray}\label{eq:phiint}
\phi(x)&=& -\varphi(x) + PV \int_0^{v_F^{-2}}d\rho\,  \big(\rho^2-\frac{3+v_F^2}{2}\rho + 1\big)\times \frac{1}{1-\rho}\nonumber\\
&\times &\big[\cos(2  x\sqrt{1-\rho})\theta(1-\rho)+ e^{-2  x \sqrt{\rho-1}}\theta(\rho - 1)\big]\nonumber\\
\equiv -\varphi(x)+\chi(x)\, .
\end{eqnarray}

The function $\chi(x)$ can be computed analytically and can be written in terms of cosine integral and exponential integral functions. We omit this long expression here, but quote its main properties. For small values of $x$, we have $\chi(0)\simeq -1/(2\, v_F^4)$. This can be explicitly checked by setting $x=0$ in the integral of Eq.\eqref{eq:phiint}. On the other hand, for large values of $x$ it oscillates with an amplitude $O(x^{-1})$.

In Fig. 4 and Fig. 5 we plot  $\phi$ and $x^5\phi$ as  functions of $x =  a \big(|\nu| - \Omega\big)$. There is an important difference between the results for parallel and perpendicular motions. At short distances from the plane, in the case of parallel motion the effect of the graphene is to decrease the probability of emission, while for perpendicular motion the 
probability is enhanced. Note that, at short distances, $|\chi(x)|\ll \varphi(x)$, and therefore $\phi(x)\simeq - \varphi(x)$, as is clear from 
Figs. 3 and 5. On the other hand, at large distances there is a partial cancellation between the $O(x^{-1})$ oscillations of $\varphi(x)$ and 
$\chi(x)$, and therefore $\phi(x)$ oscillates with a smaller amplitude,
that we found to be of order $O(x^{-2})$.

\begin{figure}
\includegraphics[width=9cm]{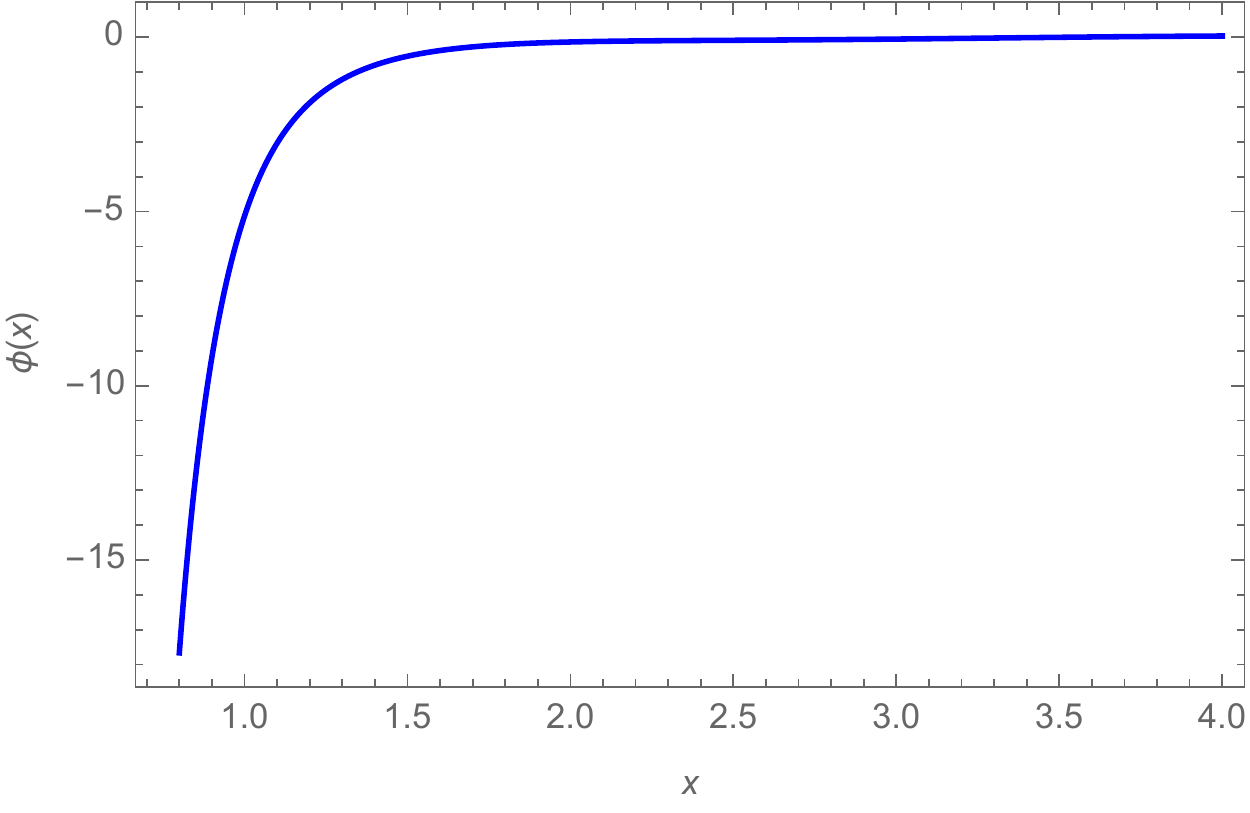}
\caption{Dimensionless function $\phi$ as a function of $x =  a \big(|\nu| - \Omega\big)$ (also dimensionless). }  \label{fig4}
\end{figure}

\begin{figure}
\includegraphics[width=9cm]{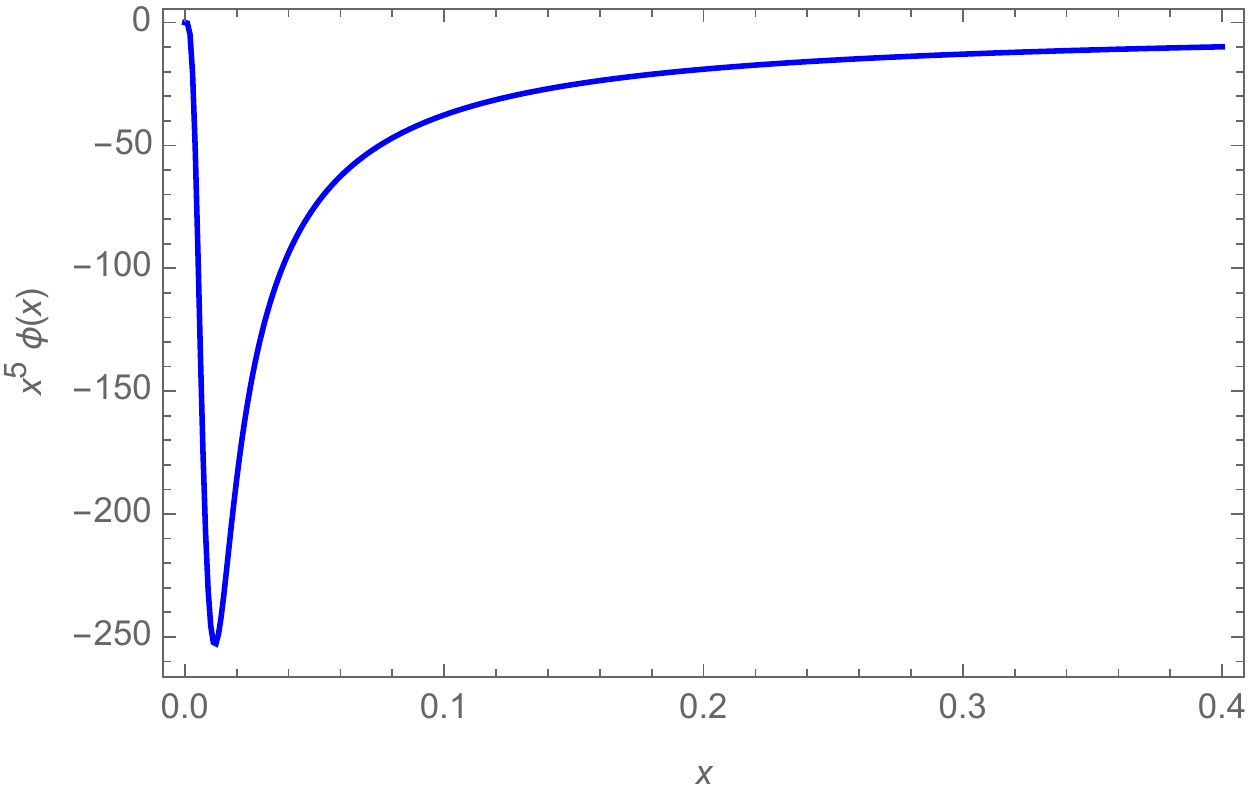}
\caption{Small $x$ behaviour of the product $x^5 \phi(x)$. }  \label{fig5}
\end{figure}
 
The fact that $m_\parallel(\nu)$ and $m_\perp(\nu)$ have different signs  also occurs for a moving atom in front of a 
perfect conductor. In that case, the different signs come from the evaluation of the two-point functions using the image method.

\section{The exact propagator for the gauge field in the presence of
the sheet}\label{sec:exact}
In \ref{ssec:second}, we calculated the imaginary part of the effective
action to the first order in both ${\mathcal S}_I^{(a)}$ and 
${\mathcal S}_I^{(g)}$. It is possible to derive an expression for
the effective action which is of the first order in ${\mathcal S}_I^{(a)}$
and exact in ${\mathcal S}_I^{(g)}$. Indeed, one can for example sum over
all the terms which appear when expanding the coupling to the sheet.
Equivalently, one can calculate the exact gauge field propagator in the
presence of the sheet, and subtract the free propagator (which has already
been considered). 

Both approaches amount, at the level of the expression for $B$, to an
identical expression as (\ref{eq:defb1}), but with different functions
$G_t$ and $G_l$ replacing $g_t$ and $g_l$ respectively, and defined as
follows:
\begin{equation}
	G_t \;=\; \frac{g_t}{1 + \frac{g_t}{ 2 \sqrt{ {\mathbf
	k}_\shortparallel^2 - k_0^2}}}
	\;\;,\;\;\;
	G_l \;=\; \frac{g_l}{1 + \frac{g_l}{ 2 \sqrt{ {\mathbf
	k}_\shortparallel^2 - k_0^2}}} \;.
\end{equation}
Therefore, it is possible to consider, within the same approach, some
interesing phenomena. For example, we may take the limit $\alpha_N \to
\infty$, under which both $G_t$ and $G_l$ tend to the same limit:
\begin{equation}
\alpha_N \to \infty \;\Rightarrow \; G_t, G_l  \to  \, 
	2 \sqrt{{\mathbf k}_\shortparallel^2 - k_0^2} \;.
\end{equation}
Now, recalling the functions $g_t$ and $g_l$ for graphene, we see that this
limit can be treated by using the graphene expressions, but with $v_F = 1$
and $\alpha_N = 2$.
In particular, 
\begin{align}
{\rm Im}\Big[B(\nu,{\mathbf k_\shortparallel},k^3,p^3)\Big] &=\;
		 \frac{ 2\pi e^2}{m \Omega} 
		\; \theta[|\nu| - \Omega - |{\mathbf k_\shortparallel}|] \nonumber\\
	&\times \;\frac{|{\mathbf k_\shortparallel}|^4 - [ 3 (|\nu|-
\Omega)^2 + k_3 p_3 ]|{\mathbf k_\shortparallel}|^2 + 2 (|\nu|-
	\Omega)^4 }{\big[|{\mathbf k_\shortparallel}|^2 +
k_3^2 - (|\nu|- \Omega)^2 \big]  \big[|{\mathbf k_\shortparallel}|^2 +
p_3^2 - (|\nu|- \Omega)^2 \big] (|\nu|- \Omega)} 
	 \;,
\end{align}
which is the kernel determining the dissipative effects in the case of an
atom moving in front of a perfectly conducting plane. Of course, the
velocity threshold means that now there will not be friction, since that
would require the atom to move at superluminal speeds.

For intermediate values of $\alpha_N$, we have:
\begin{align}
	G_t &=\; \alpha_N \frac{ ({\mathbf k_\shortparallel}^2 - k_0^2)
	\sqrt{v_F^2 {\mathbf k_\shortparallel}^2 - k_0^2}
	- \frac{\alpha_N}{2} 
	(v_F^2 {\mathbf k_\shortparallel}^2 - k_0^2)
	\sqrt{{\mathbf k_\shortparallel}^2 - k_0^2}}{\big(1 -
	(\frac{\alpha_N}{2})^2 v_F^2\big) {\mathbf k_\shortparallel}^2 -
	\big(1 - (\frac{\alpha_N}{2})^2\big) k_0^2} \nonumber\\
	G_l &=\; \alpha_N \frac{ ({\mathbf k_\shortparallel}^2 - k_0^2)
	\sqrt{v_F^2 {\mathbf k_\shortparallel}^2 - k_0^2}
	- \frac{\alpha_N}{2} 
	({\mathbf k_\shortparallel}^2 - k_0^2)
	\sqrt{{\mathbf k_\shortparallel}^2 - k_0^2}}{\big(v_F^2 - (\frac{\alpha_N}{2})^2\big)
	{\mathbf k_\shortparallel}^2 - \big(1 - (\frac{\alpha_N}{2})^2\big)
	k_0^2} \;.
\end{align}
This shows that in the small $\alpha_N$ limit, we recover the singularities
(cuts) on $k_0 = v_F |{\mathbf k_\shortparallel}|$ of the perturbative
calculation. Note, however, that that contribution is overcome by the cut 
$k_0 = |{\mathbf k_\shortparallel}|$ for bigger values of $\alpha_N$.  

The effects that would result from considering the exact gauge field
propagator (a resummation of the all terms involving a  coupling to the
medium) will, for graphene and other systems, be explored elsewhere.

\section{Conclusions}\label{sec:conc}
We have studied a model with an atom which moves non-relativistically,
both at constant and non-constant speeds, in the presence of a planar
graphene sheet. 
The model used for the atom is based on a dipole coupling which, as we have 
showed, yields essentially the same results when one uses  a harmonic
coupling of the electron to the nucleus or when one instead implements 
a two-level description. This result appears when one considers the
coupling of the atom to the EM field, and integrates out the electron's
degrees of freedom, to the lowest non-trivial order, which is quadratic in
the electric field. 
Note that, that order is exact in the harmonic case, but approximate for
a two-level system. 

The quantum dissipative effects have been studied by first deriving a
general expression for a kernel which determines them (the imaginary part
of the effective action) in terms of the motion of the atom. This general
expression does not rely on the approximation of small amplitude motion,
and therefore allows us to study both QF and the DCE on the same footing.

In the next step we considered and evaluated the effects for particular
states of motion: constant speed and bounded
motion. For the former we showed that QF exhibits the same threshold as for
two graphene sheets moving at a constant relative speed \cite{Belen2017}.
It is interesting to remark that there is also a threshold in QF  
for non-dispersive dielectrics \cite{Kardar}. 
Indeed, when considering half-spaces described by a real
constant dielectric function in relative motion, 
a frictional force arises between them when the velocity of moving half-spaces, in
their center of mass frame, is larger than the phase speed of
light in the medium (this is a quantum analog of the well-known classical
Cherenkov radiation). In this respect, graphene behaves as a non-dispersive
dielectric medium. For atoms moving near a metallic surface, although there is no threshold, QF is exponentially small at low velocities
\cite{Dalvit1}.

For bounded motion, we particularized to the case of oscillatory motions with small
amplitudes along directions which are either normal or parallel to the
sheet, at the second order in those amplitudes. 
Note that, at this order, the effects of those two motions superpose.
We have found that, in tune with the result for a moving atom near a perfect conductor, the effect of the graphene 
on the imaginary part of the effective action has different signs for normal and parallel motions.

We also pinpoint an effect which we have mentioned just when the
atom moves in vacuum, but which is of more general validity: when the
approximation of small amplitudes is not made for a simple harmonic motion,
the system receives excitations of not just the fundamental frequency but
also from its harmonics (of course with decreasing amplitudes); remember that the position of the atom appears in the exponent of a Fourier integral.

\section*{Appendix A: Evaluation of principal values}

In the evaluation of $m_\perp(\nu)$ presented in \ref{sub:perp}, it is necessary to compute integrals of the form
\begin{equation}
I_\pm(n)=PV\int \frac{dk^3}{2\pi}\int\frac{dp^3}{2\pi}\frac{e^{-ia(k^3+p^3)}(k_3p_3)^n}{(k_3^2\pm A^2)(p_3^2\pm A^2)}\, ,
\end{equation}
where $A^2 = |{\mathbf k_\shortparallel}^2-(|\nu|-\Omega)^2|$ and $n=1,2.$
It will be useful to consider also the case $n=0$.

The computation of $I_+(n)$ is straightforward and gives
\begin{align}
I_+(0) &=\; \frac{1}{4 A^2}e^{-2 A a}\nonumber\\
I_+(1) &=\; -\frac{1}{4 }e^{-2 A a}\nonumber\\
I_+(2) &=\; \frac{A^2}{4}e^{-2 A a}\, .
\end{align}
The principal value prescription is of course unnecessary in this case.

There is a subtlety in the evaluation of $I_-(n)$. Although at first
sight $I_- (n)$ is the product of two independent integrals, the principal
value of the product differs from the product of the principal values 
\cite{DaviesPV}. We illustrate the correct procedure for the particular
case $n=0$. Using Feynman parametrization and shifting the integration
variable $k^3\to k^3- p^3$ we have
\begin{equation}
I_-(0)=PV \int_0^1 dx \int\frac{dk^3}{2\pi}e^{-iak^3}\int\frac{dp^3}{2\pi}\frac{1}{p_3^2+ x(1-x) k_3^2-A^2}\,.
\end{equation}
Now we implement the principal value prescription as 
\begin{equation}
I_-(0)= {\rm Re}\big[\int_0^1 dx \int\frac{dk^3}{2\pi}e^{-iak^3}\int\frac{dp^3}{2\pi}\frac{1}{p_3^2+ x(1-x) k_3^2-A^2+ i\epsilon}\big]\,.
\end{equation}
The calculation is now standard, and yields:
\begin{equation}
I_-(0)=-\frac{1}{4 A^2}\cos(2 A a)\, .
\end{equation}

The integrals for $n=1,2$ can be computing using the following trick:
\begin{equation}
-\frac{dI_n(0)}{da^2}=PV\int \frac{dk^3}{2\pi}\int\frac{dp^3}{2\pi}\frac{e^{-ia(k^3+p^3)}\big(k_3^2+
p_3^2+ 2 k_3p_3\big)}{(k_3^2- A^2)(p_3^2- A^2)}\, .
\end{equation}
Writing
\begin{equation}
k_3^2+ p_3^2+ 2 k_3p_3 = k_3^2 - A^2+
p_3^2-A^2+ 2 k_3p_3 + 2 A^2\, ,
\end{equation}
one can easily see that only the last two terms in the integral are not vanishing for $a\neq 0$. Therefore
\begin{equation}
-\frac{dI_n(0)}{da^2}= 2 A^2I_-(0)+ 2I_-(1)\Longrightarrow  I_-(1) = A^2 I_-(0)\, .
\end{equation}

Using a similar argument that involves the fourth derivative of $I_-(0)$
one can show that 
\begin{equation}
I_-(2) = A^4 I_-(0)\, .
\end{equation}
From these results, it is rather straightforward to obtain
Eq.\eqref{eq:mperpfin} for $m_\perp(\nu)$.

\section* {Acknowledgments}
This research was supported by Agencia Nacional de Promoci\'on Cient\'ifica y Tecnol\'ogica
(ANPCyT), Consejo Nacional de Investigaciones Cient\'ificas y T\'ecnicas (CONICET), Universidad de Buenos Aires (UBA) and
Universidad Nacional de Cuyo (UNCuyo). F.C.L. acknowledges International Centre for Theoretical Physics and the Simons Associate Programme.


\begin{thebibliography}{99}

\bibitem{books}Milonni, P.W.  {\it The Quantum Vacuum}, Academic Press, San Diego, {\bf 1994}; 
Milton, K.A. 
{\it The Casimir effect: Physical manifestations of zero-point energy,}
River Edge, USA: World Scientific, {\bf 2001};
Bordag, M.; Klimchitskaya, G.L.; Mohideen, U.; and Mostepanenko, V.M. {\it Advances in the Casimir Effect},
Oxford University Press, Oxford, {\bf 2009}.

\bibitem{graphene} Castro Neto, A.H.; Guinea, F.; Peres, N.M.R.; Novoselov, K.S.; and Geim, A.K. The electronic properties of graphene, Rev. Mod. Phys {\bf 2009}, 81,109.

\bibitem{graphprop} 
Kuzmenko, A.; Van Heumen, E.; Carbone, F.; and Van Der Marel, D.  
Universal Optical Conductance of Graphite, 
Phys. Rev. Lett  {\bf 2008}, 100, 117401;
Nair, R.R.; Blake, P.; Grigorenko, A.N.; Novoselov, K.S.; Booth, T.J.; Stauber, T.; Peres, N.M.; and Geim, A.K.  
Fine structure constant defines visual transparency of graphene,
Science {\bf 2008}, 
320, 1308.

\bibitem{graphCas}
Bordag, M.; Fialkovsky, I.V.; Gitman, D.M.;  and Vassilevich, D.V.
Casimir interaction between a perfect conductor and graphene described by the Dirac model,
Phys. Rev. B {\bf 2009}, 80, 245406.

\bibitem{graphCasT}
Bimonte, G.; Klimchitskaya, G.L.;  and Mostepanenko, V.M. 
How to observe the giant thermal effect in the Casimir force for graphene systems,
Phys. Rev. A {\bf 2017}, 96, 012517.


\bibitem{graphnonideal}
Bordag, M.; Fialkovskiy, I.V. and Vassilevich, D.V. 
Enhanced Casimir effect for doped graphene,
Phys. Rev. B {\bf 2016}, 93, 075414
[erratum: Phys. Rev. B {\bf 2017}, 95, 119905].


\bibitem{graphreview}
Klimchitskaya, G.L. and Mostepanenko, V.M. 
Casimir and Casimir-Polder Forces in Graphene Systems: Quantum Field Theoretical Description and Thermodynamics,
Universe {\bf 2020}, 6, 150. 


\bibitem{abinitio}
Bordag, M.; Klimchitskaya, G.L.; Mostepanenko, V.M.; and Petrov, V.M. 
Quantum field theoretical description for the reflectivity of graphene,
Phys. Rev. D {\bf 2015}, 91, 045037
[erratum: Phys. Rev. D {\bf 2016}, 93, 089907].


\bibitem{reviewDCE}
Dodonov, V.V. 
Current status of the dynamical Casimir effect,
Phys. Scripta {\bf 2010}, 82, 038105; 
Dalvit, D.A.R.;  Maia Neto, P.A.; and Mazzitelli, F.D. 
Fluctuations, dissipation and the dynamical Casimir effect,
Lect. Notes Phys. {\bf 2011}, 834, 419-457;
Nation, P.D.; Johansson, J.R.; Blencowe, M.P.; and Nori, F. 
Stimulating Uncertainty: Amplifying the Quantum Vacuum with Superconducting Circuits,
Rev. Mod. Phys. {\bf 2012}, 84, 1; 
Dodonov, V.V. 
Fifty Years of the Dynamical Casimir Effect,
MDPI Physics {\bf 2020}, 2, 67-104.

\bibitem{qfriction} Volokitin, A. and Persson, B.N. Near-field radiative heat transfer and non-contact friction, Reviews of Modern Physics {\bf 2007}, {\bf 79}, 1291; 
Pendry, J. Shearing the vacuum - quantum friction,
Journal of Physics: Condensed Matter 
{\bf 1997}, 9 10301; Pendry, J. Quantum friction - fact or fiction?,  New Journal of Physics {\bf 2010}, 12, 033028;
Hoye, S.; Brevik, I.; and Milton, K.A.  Casimir friction between polarizable particle and half-space with radiation damping at zero temperature,  J. Phys. A {\bf 2015}, 48, 365004.

\bibitem{interplay}
Scheel, S.; and Buhmann, S.Y., Casimir-Polder forces on moving atoms,   Phys. Rev. A {\bf 2009}, 80, 042902;
Scheel, S.; and Buhmann, S.Y., Path decoherence of charged and neutral particles near surfaces,  Phys. Rev. A {\bf 2010} 85, 030101;
Impens, F.;  Behunin, R.O.; Ttira, C.C.;  Neto, P.A.M., Non-local double-path Casimir phase in atom interferometers,  Eur. Phys. Lett. {\bf 2013} 101, 60006;
Intravaia, F; et al,  Friction forces on atoms after acceleration,  J. Phys.: Condens. Matter {\bf 2015},  27 214020.


\bibitem{Moore} Moore, G.T. Quantum theory of the electromagnetic field in a variable-length one-dimensional cavity , J. Math. Phys. {\bf 1970}, 11, 2679.

\bibitem{Fulling}
Davies, P.C.W. and Fulling, S.A.
Radiation from Moving Mirrors and from Black Holes,
Proc. Roy. Soc. Lond. A {\bf 1977}, 356, 237-257.

\bibitem{Paulo2018} de Melo e Souza, R.; Impens, F.; and Maia Neto, P.A. 
Microscopic dynamical Casimir effect, 
Phys. Rev. A {\bf 2018}, 97, 032514.

\bibitem{Belen2019}
Far\'\i as, M.B.; Fosco, C.D.; Lombardo, F.C.; and Mazzitelli, F.D. 
Motion induced radiation and quantum friction for a moving atom,
Phys. Rev. D {\bf 2019}, 100, 036013.

\bibitem{dcemat} Yablonovitch, E. Accelerating reference frame for electromagnetic waves in a rapidly growing plasma: Unruh-Davies-Fulling-DeWitt radiation and the nonadiabatic Casimir effect,  Phys. Rev. Lett. {\bf 1989}, 62, 1742;
Agnesi, A.; Braggio, C.; Bressi, G.;  Carugno, G.; Della
Valle, F.; Galeazzi, G.; Messineo, G.; Pirzio,  F.; Reali, G.; and
Ruoso, G. MIR: An experiment for the measurement of the
dynamical Casimir effect, Journal of Physics: Conference Series {\bf 2009}, 161,
012028.

\bibitem{dceexp}Wilson, C.M.; Johansson, G.; Pourkabirian, A.; Simoen, M.; 
Johansson, J.R.; Duty, T.; Nori, F.;  and Delsing, P. Observation of the dynamical Casimir effect in a superconducting circuit, Nature {\bf 2011}, 
479, 376.

\bibitem{Lomb2020}Far\'ias, M.B.; Lombardo, F.C.; Soba, A.; Villar, P.I.; and Decca, R.S. Towards detecting traces of non-contact 
quantum friction in the corrections of the accumulated geometric phase, 
npj Quantum Information {\bf 2020}, 6 (1), 1-7.

\bibitem{Lomb2021} Viotti, L.; Lombardo, F.C.; and Villar, P.I. 
Enhanced decoherence for a neutral particle sliding on a metallic surface in vacuum, 
Phys. Rev. A {\bf 2021}, 103, 032809.

\bibitem{qute21} Lombardo, F.C.; Decca, R.S.; Viotti, L.; and Villar, P.I.  
Detectable Signature of Quantum Friction on a Sliding Particle in Vacuum,
Adv. Quantum Tech. {\bf 2021}, 2000155, 1-9.


\bibitem{Belen2017}
Far\'\i as, M.B.; Fosco, C.D.; Lombardo, F.C.;  and Mazzitelli, F.D. 
Quantum friction between graphene sheets, 
Phys. Rev. D {\bf 2017}, 95, 065012

\bibitem{spontemis}Barnett, S.M.; Huttner, B.;  and Loudon, R.  
Spontaneous emission in absorbing dielectric media,
Phys. Rev. Lett. {\bf 1992}, 68, 3698; Barnett S.M.; Huttner, B.; Loudon, R.;  and Matloob, M.  
Decay of excited atoms in absorbing dielectrics, 
J. Phys. B: At. Mol. Opt. Phys. {\bf 1996}, 
29,  3763.

\bibitem{FLM2007}Fosco, C.D.; Lombardo, F.C.; and Mazzitelli, F.D. 
Quantum dissipative effects in moving mirrors: A Functional approach,
Phys. Rev. D {\bf 2007}, 76, 085007.


\bibitem{Giraldo}
Fosco, C.D.; Giraldo, A.;  and Mazzitelli, F.D. Dynamical Casimir effect for semitransparent mirrors,
Phys. Rev. D {\bf 2017}, 96, 045004.

\bibitem{Kardar}
Maghrebi, M.F.; Golestanian, R.;  and Kardar, M. 
Quantum Cherenkov Radiation and Non-contact Friction,
Phys. Rev. A {\bf 2013}, 88, 042509.

\bibitem{Dalvit1} Intravaia, F.; Behunin, R.O.; Henkel, C.; Busch, K.; and Dalvit, D.A.R. Non-Markovianity in atom-surface dispersion forces, 
Phys. Rev.  A {\bf 2016}, 94, 042114.

\bibitem{DaviesPV} Davies, K.T.R.; Glasser, M.L.; Protopescu, V.; and Tabakin, F.  
The mathematics of principal value integrals and applications to nuclear physics, transport theory, and condensed matter physics, 
Mathematical Models and Methods in Applied Sciences {\bf 1996}, 06, 833.

\end{thebibliography}
\end{document}